\documentclass[12pt]{article}
\usepackage[letterpaper]{geometry}
\usepackage{amsmath,amssymb,amsthm,bm,graphicx,float,array,multirow,multicol,rotfloat,caption,subcaption,hyperref,cleveref,enumerate,mathdots,adjustbox,booktabs,parskip,mathtools,tikz,tikz-cd,pdflscape,csquotes,lscape,rotating,empheq,mathrsfs,longtable}
\usepackage[para]{threeparttable}
\usepackage[all]{xy}
\usepackage[normalem]{ulem}
\usepackage[aligntableaux=center]{ytableau}
\usepackage[numbers,sort&compress]{natbib}

\usepackage{xeCJK}
\usepackage{fontspec}

\usepackage{comment}

\numberwithin{equation}{section}
\numberwithin{figure}{section}
\allowdisplaybreaks
\makeatletter
\usetikzlibrary{arrows,shapes.misc,positioning,decorations.pathmorphing,decorations.markings,decorations.pathreplacing,matrix,patterns,backgrounds}

\definecolor{gold}{RGB}{255,215,0}
\definecolor{purple}{RGB}{160,32,240}
\tikzset{
scale cd/.style={every label/.append style={scale=#1}, cells={nodes={scale=#1}}},
gauge/.style={rounded rectangle, draw=black!100, thick, minimum size=5mm}, 
gaugeD/.style={rounded rectangle, draw=black!100,double,thick,minimum size=5mm},  
empty/.style={rounded rectangle, draw=white!100, thick, minimum size=5mm}, 
flavor/.style={rectangle, draw=black!100, thick, minimum size=5mm},
flavorD/.style={rectangle, draw=black!100, double,thick, minimum size=5mm},
node/.style={circle, thick, draw=black!100,fill=white!100,  minimum size=2mm, inner sep=0pt},
sqnode/.style={rectangle
, thick, draw=black!100,fill=white!100,  minimum size=2mm, inner sep=0pt
},
sonode/.style={circle, thick, draw=black!100,fill=red!100,  minimum size=3mm, inner sep=0pt},
spnode/.style={circle, thick, draw=black!100,fill=blue!100,  minimum size=3mm, inner sep=0pt},
fnode/.style={rectangle, thick, draw=black!100,fill=white!100,  minimum size=3mm, inner sep=0pt},
tnode/.style={rounded rectangle, outer sep=0pt, thick, minimum size=5mm},
brace/.style={decoration={brace, mirror},decorate}
}
\DeclarePairedDelimiter\floor{\lfloor}{\rfloor}

\newcommand\notsoscript{\@setfontsize\notsoscript{9}{7}}

\theoremstyle{plain}
\newtheorem*{thm*}{Theorem}

\theoremstyle{definition}

\newtheorem*{defn*}{Definition}

\makeatother

\def\scheme/{our affine scheme} 

\graphicspath{{figures/}}

\begin{document}

\begin{titlepage}
\vspace*{-3cm} 
\begin{flushright}
{\tt DESY-26-087}\\
{\tt CALT-TH 2026-026} \\
\end{flushright}
\begin{center}
\vspace{1.2cm}

{\LARGE\bfseries The index from a nonpoint and RG flows}
\vspace{.9cm}

{\large
Monica Jinwoo Kang,$^{1}$ Craig Lawrie,$^{2}$ and Jaewon Song$^{3,4}$\\}
\vspace{.7cm}
{$^1$ Mitchell Institute for Fundamental Physics and Astronomy, Texas A\&M University,\\
College Station, TX 77843, U.S.A.}\par
\vspace{.1cm}
{$^2$ Deutsches Elektronen-Synchrotron DESY,
Notkestr.~85, 22607 Hamburg, Germany}\par
\vspace{.1cm}
{$^3$ Department of Physics, Korea Advanced Institute of Science and Technology\\
Daejeon 34141, Republic of Korea}\par
\vspace{.1cm}
{$^4$ Walter Burke Institute for Theoretical Physics \& Leinweber Forum for Theoretical Physics\\ California Institute of Technology, 
Pasadena, CA 91125, USA}\par
\vspace{.3cm}

\scalebox{.9}{\tt monicak@tamu.edu, craig.lawrie1729@gmail.com, jaewon.song@kaist.ac.kr}\par
\vspace{0.7cm}
\textbf{Abstract}
\end{center}
{
We extend the algebro-geometric construction of bifiltered affine schemes that encode both the Higgs branch and the Macdonald index of 4d $\mathcal{N}=2$ superconformal field theories to theories with non-trivial Higgs branches.
For the Argyres--Douglas theories $\mathcal{D}_p(SU(N))$, we find that a universal potential, given by a function in the Casimir invariants of $\mathfrak{sl}(N)$, determines the scheme. 
The scheme is given by a `fattened' nilpotent orbit closure, encoding the OPE decoupling structure of the 4d theory. 
A procedure we call syzygy maximization fixes the precise potential in a broad range of examples, and the resulting schemes reproduce the 4d data. The underlying graded ring gives conjectural presentations of Zhu's $C_2$-algebra of the associated vertex operator algebra, while the bifiltration refines Arakawa's associated scheme by retaining the R-symmetry information. 
We find that Higgs branch renormalization group flows are realized by pulling back the same universal potential to the corresponding Slodowy slices, including flows for which the infrared Higgs branch is a point.  
In particular, our construction provides a systematic method for determining the Macdonald indices of the $\mathcal{D}_p(SU(N), O)$ theories, for which no general expressions were previously known.

\vspace{0.8cm}
\small
\begin{center}
``以指喻指之非指~~不若以非指喻指之非指也'' -- 莊子
\\ \vspace{0.2cm}
``Using a point to show that a point is not a point is not as good as \phantom{aaaaaaaaaa} \newline \phantom{aaaaaaaaaa} using a nonpoint to show that a point is not a point.'' -- Zhuangzi
\end{center}
\normalsize
}
\vfill 
\end{titlepage}

\tableofcontents
\newpage

\section{Introduction}\label{sec:intro}

In our recent work \cite{Kang:2025zub,Kang:2026nge,SUPERPOINT}, we have proposed that one can associate to a 4d $\mathcal{N}=2$ SCFT a bifiltered commutative ring $R$, such that $R$ encodes both the Higgs branch of the SCFT and the Macdonald index of 1/4-BPS operators.\footnote{See also \cite{Bhargava:2023hsc,Andrews:2025krn} for related ideas.} The ring is proposed to be associated to the OPE decoupling of operators, as studied in \cite{Agarwal:2018zqi,Liendo:2015ofa}. The Higgs branch is recovered by taking the reduced spectrum of the ring:
\begin{equation}\label{eqn:bigHIGGS}
    \mathcal{M}_{\mathcal{H}} = \operatorname{Spec}(R)_\text{red} \,.
\end{equation}
As we will clarify later, this ring $R$ is a refined/extended version of Zhu's $C_2$-algebra, $R_\mathbb{V}$, for the vertex operator algebra $\mathbb{V}$ associated to the 4d $\mathcal{N}=2$ SCFT \cite{Beem:2013sza}, and thus it is related to the coordinate ring of Arakawa's associated scheme \cite{arakawa2015associated}, which itself is conjectured to reduce to the Higgs branch \cite{Beem:2017ooy, Song:2017oew}. 
The Macdonald index, which we explain how to recover from $R$ anon, is defined as \cite{Gadde:2011uv}
\begin{align}
    I_\text{Mac}(q, T) = \operatorname{Tr} \, (-1)^F q^{\Delta - R} T^{R - r} \,,
\end{align}
where the trace is taken over $1/4$-BPS operators satisfying $\Delta - 2j_1 - 2R - r = 0$ and $j_1 - j_2 + r = 0$. Here, $\Delta$ is the conformal dimension, $(j_1, j_2)$ are the Lorentz spins, and $R$ and $r$ are the charges under the $\mathfrak{su}(2)_R$ and $\mathfrak{u}(1)_r$ R-symmetries, respectively.

The ring $R$ is assumed to be finitely-generated, in which case it can   be written (non-uniquely) as a quotient ring
\begin{equation}
    R = S \, / \, I \,,
\end{equation}
where $S$ is a polynomial ring, and $I$ is an ideal belonging to $S$. In the cases that we consider, the bifiltered structure on $R$ comes from a bigrading on $S$, which we refer to as the $q$-grading and the $T$-grading; the ideal, $I$, is generated by polynomials which are homogeneous under the $q$-grading, but not necessarily homogeneous under the $T$-grading. Thus, the quotient, $R$, has an induced bifiltered structure. Our proposal \cite{Kang:2025zub,Kang:2026nge} is then that the ``Hilbert series of the arc space'' of $R$ is identical to the Macdonald index:\footnote{We have extended our proposal to rings that are $\mathbb{Z}_2$-graded-commutative, i.e., that have fermionic generators, in \cite{SUPERPOINT}, but this generalization is not necessary for the current work.}
\begin{equation}\label{eqn:bigMAC}
    I_\text{Mac}(q, T) = \operatorname{HS}_{q,q,T}\left(\operatorname{gr}\left(J_\infty(R)\right)\right) \,.
\end{equation}
The terminology, especially on the RHS, is explained in detail in \cite{Kang:2025zub,Kang:2026nge}. 
Schematically, the arc space $J_\infty (R)$ introduces effective holomorphic coordinates that map to space-time, enabling us to introduce a jet-grading, and the graded vector space $\operatorname{gr}(V)$ made out of $V=J_\infty(R)$ turns a filtered algebra into a graded algebra. Therefore, we get a triply-graded vector space by this procedure, which captures space-time derivatives and $T$-grading.\footnote{The second part is universal and is essentially the same procedure as the one introduced in \cite{Song:2016yfd} to recover Macdonald grading from the VOA associated to 4d SCFT. See also \cite{Fluder:2017oxm, Agarwal:2018zqi,Bonetti:2018fqz, Beem:2019tfp, Xie:2019zlb, Watanabe:2019ssf, Agarwal:2021oyl, ArabiArdehali:2025fad}. } 
The Schur index, which is the $T = 1$ limit of the Macdonald index, is obtained by forgetting about the $T$-grading on $S$, in which case $R$ is a graded ring, rather than a bifiltered ring. We write $R_S$ to denote the graded ring obtained from forgetting about this extra structure in the bifiltration. Thus,
\begin{equation}\label{eqn:bigSCHUR}
    I_\text{Schur}(q) = \operatorname{HS}_{q,q}(J_\infty(R_S)) \,.
\end{equation}

In \cite{Kang:2025zub,Kang:2026nge}, the principal focus was on theories with a trivial Higgs branch. In particular, we studied the Argyres--Douglas theories $(A_{k-1}, A_{N-1})$ \cite{Argyres:1995jj, Argyres:1995xn, Xie:2012hs} with $\gcd(k,N) = 1$. In such cases, we found that the bifiltered ring can be obtained as follows. We start with the closed-form expression for the Schur indices of such $(A_{k-1}, A_{N-1})$ theories \cite{Cordova:2015nma, Song:2015wta, Song:2017oew}:
\begin{equation}\label{eqn:SchurPE}
   I_\text{Schur}(q) = \operatorname{PE} \left[\frac{q^2 + \cdots + q^k - q^{N + 1} - \cdots - q^{k + N - 1}}{(1-q)(1-q^{k+N})} \right] \,,
\end{equation}
where $\operatorname{PE}$ denotes the standard plethystic exponential. We seek a ring $R = S \, / \, I$ which reproduces this expression via equation \eqref{eqn:bigSCHUR}.
Since we are considering the Schur index, let us forget about the $T$-grading of $S$, in which case the resulting $R_S$ is a graded ring, with the grading inherited from the $q$-grading of $S$, under which $I$ is homogeneous. The coordinate ring of the arc space, $J_\infty(R_S) = S_\infty \, / \, I_\infty$, is a bigraded ring, where one of the gradings is induced by the $q$-grading, and the other is the jet-grading.\footnote{We sometimes refer to the jet-grading as the $p$-grading.} 

The Hilbert series of any (bi)graded ring can be determined from the minimal free resolution of the ring, which is unique up to isomorphism of complexes.\footnote{For a review of the relevant concepts in commutative algebra, we refer to \cite{MR1322960}, or the upcoming \cite{BETTI}.} The minimal free resolution is an exact sequence
\begin{equation}
    \cdots \rightarrow F_2 \xrightarrow{d_2} F_1 \xrightarrow{d_1} S_\infty \xrightarrow{d_0} J_\infty(R_S) \rightarrow 0 \,.
\end{equation}
Here the $F_i$ are bigraded free $S_\infty$-modules:
\begin{equation}
    F_i = \bigoplus_{(a,b) \in \mathbb{Z}^2}S_\infty(-a,-b)^{\beta_{i,(a,b)}} \,,
\end{equation}
where $S_\infty(-a,-b)$ denotes a free rank-one copy of $S_\infty$ whose basis element has bidegree $(a,b)$.
The exponents, the $\beta_{i,(a,b)}$, are called the algebraic Betti numbers, and they count the number of minimal generators of the $i$th syzygy module of $J_\infty(R_S)$ of given bidegree; they are intrinsic properties of the ring. In terms of these algebraic Betti numbers, the Hilbert series can be written as
\begin{equation}
    \operatorname{HS}_{p,q}(J_\infty(R_S)) = \operatorname{HS}_{p,q}(S_\infty) \, \left( 1 + \sum_{i \geq 1} (-1)^i \sum_{(a,b) \in \mathbb{Z}^2} \beta_{i,(a,b)} p^b q^a \right) \,.
\end{equation}
Matching this expression, order-by-order, to the closed-form for the Schur index in equation \eqref{eqn:SchurPE}, constrains the algebraic Betti numbers. Furthermore, given the known data of the $(A_{k-1}, A_{N-1})$ Argyres--Douglas theories, these Betti numbers uniquely fix the ring $R_S$.

In this way, we found that the following information was necessary and sufficient to completely fix the ring $R$ underlying the Macdonald index and the Higgs branch of the $(A_{k-1}, A_{N-1})$ Argyres--Douglas theories with $\gcd(k,N)=1$ \cite{Kang:2026nge}. Consider the ring 
\begin{equation}
    R = S \, / \, I \qquad \text{ such that } \qquad S = \mathbb{C}[x_{2,1}, \cdots, x_{k,k-1}]  \,,
\end{equation}
where the subscripts denote the $(q,T)$-bidegree. Let $P(x_{2,1}, \cdots, x_{k,k-1})$ be a $q$-degree $k+N+1$ polynomial in $S$, and let $I$ be an ideal of $S$ generated by relations
\begin{equation}
    I = \left( \partial_{x_{2,1}} P, \cdots, \partial_{x_{k,k-1}} P\right) \,.
\end{equation}
We refer to the polynomial $P$ as the potential.
The ideal, $I$, and thus the ring, $R$, depend heavily on the choice of potential, $P$. We can constrain $P$ such that $R$ is a complete intersection and that 
\begin{equation}\label{eqn:smax}
    \beta_{2,(k+N+2,1)}(J_1(R_S)) = 1 \,.
\end{equation}
It turns out that these conditions uniquely fix the ideal, $I$; with this particular $I$, the bifiltered ring $R = S \,/\, I$ is precisely the ring such that equations \eqref{eqn:bigHIGGS} and \eqref{eqn:bigMAC} are satisfied.\footnote{Interestingly, in \cite{Xie:2019zlb}, it was conjectured that Zhu's $C_2$-algebra \cite{MR1317233} associated to the $(A_{k-1}, A_{N-1})$ theory, which is isomorphic as a graded ring to our $R_S$ (i.e., after forgetting about the $T$-grading), is a quotient by a Jacobian ideal. In \cite{Kang:2026nge}, we realized this conjecture by providing the precise $P$ with coefficients.} We refer to the condition in equation \eqref{eqn:smax} as ``\emph{syzygy maximization}'' since, once all the other conditions are satisfied, this Betti number is generically zero, and is non-zero at a unique point in the moduli space.

In this paper, we extend this analysis to theories where the Higgs branch is \emph{not} a point. The extension that we discover, which applies to more general Argyres--Douglas theories, behaves in the following way. There is a bigraded polynomial ring $S$, and a potential $P$ of $q$-degree $d$ where the Jacobian ideal associated to $P$, which we denote as $I$, is such that 
\begin{equation}
    \mathcal{M}_{\mathcal{H}} = \operatorname{Spec}( R = S \, / \, I )_\text{red} \,.
\end{equation}
This does not uniquely fix the ideal $I$, however, we find that if we impose a condition on the syzygies of the first jet scheme:
\begin{equation}\label{eqn:smaxgen}
    \beta_{2,(\ell, 1)}(J_1(R_S)) = 1 \,,
\end{equation}
then $I$ is uniquely fixed and the Hilbert series of the arc space of $R$ reproduces the Macdonald index. The $q$-degrees of the potential and the syzygy condition, $d$ and $\ell$, are dependent on the SCFT under consideration. In many cases, closed-form expressions for the Macdonald index of the 4d $\mathcal{N}=2$ SCFTs are not known,\footnote{Macdonald indices are known for $(A_1, A_n)$, $(A_1, D_n)$ theories, which can be computed using $\mathcal{N}=1$ Lagrangian description \cite{Maruyoshi:2016tqk, Maruyoshi:2016aim, Agarwal:2016pjo, Cho:2026tqf} or the TQFT description \cite{Buican:2015tda, Song:2015wta}. More recently, it was found that it can be obtained from the BPS spectrum on the Coulomb branch \cite{Andrews:2025krn, Kim:2025klh}.} especially to high orders. In these cases, we can match to particular, known, limits of the Macdonald index \cite{Gadde:2011uv}, which gives us confidence that the Hilbert series of the arc space of the bifiltered ring we identify is indeed giving the correct result. The simplest limit is that of the Schur limit
\begin{equation}
    I_\text{Schur}(q)\,\, =\,\, \lim_{\,T \,\rightarrow\, 1\,}\, I_\text{Mac}(q, T) \,,
\end{equation}
which corresponds, as discussed, to forgetting about the part of the bifiltration induced by the $T$-grading. Another limit which we match is the Hall--Littlewood limit:
\begin{equation}
    I_\text{HL}(\tau)\,\, =\,\, \lim_{\,q \,\rightarrow\, 0\,}\, I_\text{Mac}\left(q, \frac{\tau}{q}\right) \,.
\end{equation}
This limit keeps only the monomials appearing in the Macdonald index of the form $q^n T^n$, for any $n$, which may be half-integer. The Hall--Littlewood limit provides a strong check of the $T$-grading (which is invisible to the Schur limit), when the Macdonald index itself is not known. 

The quintessential difference between the SCFTs where the Higgs branch is a point, studied in \cite{Kang:2026nge}, and the theories studied here with a non-trivial Higgs branch, is that the latter admit Higgs branch renormalization group flows. The Higgs branch of a 4d $\mathcal{N}=2$ SCFT is both hyperK\"ahler \cite{cmp/1104116624} and a symplectic singularity \cite{Beauville_2000,Kaledin_2006}. In particular, the foliation structure of the symplectic singularity captures Higgs branch renormalization group flows between SCFTs \cite{Bourget:2019aer}.

Let us suppose that we have a theory $\mathcal{T}_{\operatorname{UV}}$, with a non-trivial Higgs branch, and another SCFT $\mathcal{T}_{\operatorname{IR}}$, which $\mathcal{T}_{\operatorname{UV}}$ flows to under a Higgs branch renormalization group (RG) flow. Furthermore, suppose we have a bifiltered ring $R_{\operatorname{UV}}$ which underlies the Macdonald index of $\mathcal{T}_{\operatorname{UV}}$. It is natural to ask if we can then determine the bifiltered ring $R_{\operatorname{IR}}$ associated to $\mathcal{T}_{\operatorname{IR}}$, starting from $R_{\operatorname{UV}}$. That is, we would like to understand if the question-marked arrow exists, and, if it does, how it acts on the bifiltered rings:
\begin{equation}
    \begin{tikzcd}[column sep=4cm, row sep=1cm]
        {R}_{\operatorname{UV}}\arrow{r}{\text{Macdonald reconstruction}} \arrow[d,swap]{}{?} & \mathcal{T}_{\operatorname{UV}}\arrow[d]{}{\text{HB RG flow}}\\
        {R}_{\operatorname{IR}}\arrow{r}{\text{Macdonald reconstruction}} & \mathcal{T}_{\operatorname{IR}}
    \end{tikzcd} \,.
\end{equation}
Where we can determine both $R_{\operatorname{UV}}$ and $R_{\operatorname{IR}}$, we find that they are related by a simple geometric action on the potentials, $P_{\operatorname{UV}}$ and $P_{\operatorname{IR}}$, that construct the relevant Jacobian ideals.

\paragraph{On the connection to VOA:}
The association of a bifiltered ring to a 4d $\mathcal{N}=2$ SCFT is reminiscent of the association of the vertex operator algebra (VOA) to each of the same class of SCFTs \cite{Beem:2013sza}. This VOA encodes the closed subsector of the 4d theory formed by the $1/4$-BPS Schur operators. For a generic 4d $\mathcal{N}=2$ SCFT, $\mathcal{T}$, we refer to the associated VOA as $\mathbb{V}_{\mathcal{T}}$. 
To each $\mathbb{V}_{\mathcal{T}}$, there is an associated commutative ring, $\mathcal{R}_{\mathbb{V}_{\mathcal{T}}}$, which is known as Zhu's $C_2$-algebra \cite{MR1317233}. This ring is defined as a quotient of the VOA as follows:\footnote{See \cite{AM} for standard definitions relevant for our discussion of vertex operator algebras.}
\begin{equation}
    C_2(\mathbb{V}_{\mathcal{T}}) = \operatorname{Span}\left( a_{(-h_a - 1)}b \, | \, a, b \in \mathbb{V}_{\mathcal{T}} \right) \,, \qquad \mathcal{R}_{\mathbb{V}_{\mathcal{T}}} = \mathbb{V}_{\mathcal{T}} \, / \, C_2(\mathbb{V}_{\mathcal{T}}) \,,
\end{equation}
where $h_a$ is the conformal dimension of the operator $a$.\footnote{In the math literature, $a_{-1}$ is typically used instead of $a_{-h_a}$, for the generator of the primary state. We use the `physics convention' here to highlight the distinct conformal dimensions given to each generator.} The subspace $C_2(\mathbb{V}_{\mathcal{T}})$ is spanned by states containing at least one derivative, so that passing to $\mathcal{R}_{\mathbb{V}_{\mathcal{T}}}$ removes derivative data.
Arakawa's associated variety is defined as $X_{\mathbb{V}_\mathcal{T}} = \operatorname{Spec}(\mathcal{R}_{\mathbb{V}_\mathcal{T}})_{\textrm{red}}$, which is conjectured to reproduce the Higgs branch of $\mathcal{T}$. 

Zhu's $C_2$-algebra $R_{\mathbb{V}_{\mathcal{T}}}$ is a graded ring, where the grading is inherited from the conformal weight of the operators belonging to the VOA. In fact, there exists a canonical \emph{descending} filtration on $\mathbb{V}_{\mathcal{T}}$, which is known as the Li filtration \cite{MR2172688}. The filtration is defined via
\begin{equation}
    F^p(\mathbb{V}_{\mathcal{T}}) = \operatorname{Span}\left( a_{(-n_1-h_{a_1})}^1 \, \cdots \, a_{(-n_r- h_{a_r})}^r \right) \,,
\end{equation}
where $a^j$ belong to the VOA and the $n_j$ are non-negative integers such that $\sum_j n_j \geq p$. There are inclusions $F^p(\mathbb{V}_{\mathcal{T}}) \supseteq F^{p+1}(\mathbb{V}_{\mathcal{T}})$, and thus we have a filtration. The associated graded ring to the Li filtration
\begin{equation}
    \operatorname{gr}_{F}(\mathbb{V}_{\mathcal{T}}) = \bigoplus_p F^p(\mathbb{V}_{\mathcal{T}}) \, / \, F^{p+1}(\mathbb{V}_{\mathcal{T}}) \,,
\end{equation}
is a commutative differential algebra generated by $R_{\mathbb{V}_{\mathcal{T}}}$ \cite{MR2048777}. Furthermore, the character of $\mathbb{V}_{\mathcal{T}}$ (which is proportional to the Schur index of the 4d theory) can be written as the Hilbert series of the associated graded ring:
\begin{equation}\label{eqn:charHS}
    \operatorname{ch}_{\mathbb{V}_{\mathcal{T}}}(q) = q^{-c/24} \, \operatorname{HS}_q(\operatorname{gr}_{F}(\mathbb{V}_{\mathcal{T}})) \,,
\end{equation}
where we note that the Hilbert series has only a single fugacity, associated to the grading of $R_{\mathbb{V}_{\mathcal{T}}}$ inherited from the conformal weight. 
Let us emphasize that for the \emph{bifiltered} ring $R$ we associate to a 4d theory, it has an \emph{ascending} filtration, opposite to the case of Li filtration. This filtration gives the desired $T$-grading upon taking the associated graded.  

Since the associated graded ring is generated as a commutative differential algebra by $R_{\mathbb{V}_{\mathcal{T}}}$, 
there exists a surjective homomorphism of differential algebras
\begin{equation}\label{eqn:CF}
    \Phi :\, J_\infty(\mathcal{R}_{\mathbb{V}_{\mathcal{T}}}) \,\rightarrow\, \,\operatorname{gr}_{F}(\mathbb{V}_\mathcal{T}) \,.
\end{equation}
If this map is actually an isomorphism, $\mathbb{V}_\mathcal{T}$ is referred to as ``classically free'' \cite{MR4321999,MR4779563}; furthermore, there is a conjecture that any VOA associated to a 4d $\mathcal{N}=2$ SCFT via the SCFT/VOA correspondence is classically free \cite{RastelliTalk1,RastelliTalk2}.\footnote{Let us comment that there exists at least one example of a 4d $\mathcal{N}=2$ SCFT, $\mathcal{T}$, where $\mathbb{V}_\mathcal{T}$ is not classically-free, and thus the Schur index is not recovered from the Hilbert series of the arc space of  $R_{\mathbb{V}_{\mathcal{T}}}$ as in equation \eqref{eqn:squirrel}. However, we find that there still exists an $R_S \neq R_{\mathbb{V}_{\mathcal{T}}}$ which reproduces the Schur index via a generalization of equation \eqref{eqn:hamster}. We discuss such examples in detail in \cite{SUPERPOINT}.} If the VOA is classically free, then equation \eqref{eqn:charHS} implies that
\begin{equation}\label{eqn:squirrel}
    I_\text{Schur}(q) = \operatorname{HS}_{q}(J_\infty(\mathcal{R}_{\mathbb{V}_{\mathcal{T}}})) \,.
\end{equation}
On the other hand, we also have
\begin{equation}\label{eqn:hamster}
    I_\text{Schur}(q) = \operatorname{HS}_{q, q}(J_\infty(R_S)) \,.
\end{equation}
where $R_S$ is the underlying $q$-graded ring of our bifiltered ring $R$ obtained by `forgetting' the $T$-filtration structure. 
In both this paper and our previous work \cite{Kang:2025zub,Kang:2026nge}, we determined a bifiltered ring $R$ such that we reproduce the Macdonald (and thus the Schur) index via the arc space construction as in equation \eqref{eqn:bigMAC}.

Based on the existence of such a ring $R$, we conjecture that the VOAs associated to the SCFTs we study in this paper (namely the ones associated to a class of Argyres--Douglas theories and their deformations) are such that $R_{\mathbb{V}_{\mathcal{T}}} \cong R_S$ and classically free. Notice that a priori these two objects are determined differently. Namely for $R$ and $R_S$, it was crucial to apply the syzygy maximization as in equation \eqref{eqn:smaxgen}, whereas $\mathcal{R}_{\mathbb{V}_\mathcal{T}}$ was obtained from the VOA $\mathbb{V}_\mathcal{T}$. 
This raises a natural question: what are the relevant properties of a VOA (that can be associated to a 4d $\mathcal{N}=2$ SCFT) that guarantee that Zhu's $C_2$-algebra is syzygy-maximizing, as in equation \eqref{eqn:smaxgen}? Conversely, does syzygy maximization allow the uplift of $R$ to $\mathbb{V}$? In fact, there are many intriguing connections between our work and the interpretation in the associated VOAs, but we leave a deeper elucidation to future work.

\begin{table}[t]
    \centering
    \begin{threeparttable}
    \begin{tabular}{ccc}
        \toprule
         SCFT ($\mathcal{T}$) & Range & Higgs Branch \\\midrule
         $(A_1, A_{2n-1})$ & $n = 2, \cdots, 10$ & $\mathbb{C}^2/\mathbb{Z}_n$ \\ \midrule
         $\mathcal{D}_p(SU(3), [2,1])$ & $p = 4, 5, 7, 8, 10, 11$ & $\mathbb{C}^2/\mathbb{Z}_3$ \\\midrule
         $\mathcal{D}_p(SU(3))$ & \begin{tabular}{c}
             $p=2$\\
             $p = 4, 5, 7, 8, 10, 11$
         \end{tabular} & \begin{tabular}{c}             
             $\overline{\mathcal{O}}_{[2,1]}^{\,\mathfrak{su}(3)}$ \\
             $\overline{\mathcal{O}}_{[3]}^{\,\mathfrak{su}(3)}$
         \end{tabular}\\\midrule
         $\mathcal{D}_7(SU(4), [3,1])$ & -- & $\mathbb{C}^2/\mathbb{Z}_4$ \\\midrule
         $\mathcal{D}_p(SU(4))$ & \begin{tabular}{c}
             $p=3$\\
             $p = 5,7$
         \end{tabular} & \begin{tabular}{c}             
             $\overline{\mathcal{O}}_{[3,1]}^{\,\mathfrak{su}(4)}$ \\
             $\overline{\mathcal{O}}_{[4]}^{\,\mathfrak{su}(4)}$
         \end{tabular}\\\midrule
         $\mathcal{D}_p(SU(5))$ & \begin{tabular}{c}
             $p=2$ \\
             $p=3$\\
             $p = 4$ \\
             $p = 6$
         \end{tabular} & \begin{tabular}{c}             $\overline{\mathcal{O}}_{[2,2,1]}^{\,\mathfrak{su}(5)}$ \\
         $\overline{\mathcal{O}}_{[3,2]}^{\,\mathfrak{su}(5)}$ \\
             $\overline{\mathcal{O}}_{[4,1]}^{\,\mathfrak{su}(5)}$ \\
             $\overline{\mathcal{O}}_{[5]}^{\,\mathfrak{su}(5)}$
         \end{tabular} \\\bottomrule
    \end{tabular}
    \end{threeparttable}
    \caption{The Argyres--Douglas SCFTs for which we herein discover the bifiltered rings underlying the Macdonald index and Higgs branch. The Higgs branches are non-trivial and are either orbifold singularities or closures of nilpotent orbits. Here, we write the bifiltered rings discovered via explicit syzygy maximization; from the ring for $\mathcal{D}_p(SU(N))$ given here, we can determine the bifiltered ring for any $\mathcal{D}_p(SU(N), O)$ using the proposal in Section \ref{sec:conj}.}
    \label{tbl:theories}
\end{table}

\paragraph{Organization of the paper} 
The rest of this paper is organized as follows. First, in Section
\ref{sec:conj}, we provide a conjecture for the bifiltered rings associated to
a broad class of Argyres--Douglas SCFTs. Then, in Section \ref{sec:RINGS}, we
derive the bifiltered rings, using the procedure of syzygy maximization,  for a
variety of explicit examples of such Argyres--Douglas theories where the Higgs
branch is non-trivial and test the Hilbert series of their arc spaces against the (known limits of the) Macdonald index. In particular, we study the SCFTs listed in Table
\ref{tbl:theories}, where the Higgs branches are either orbifold singularities
or closures of nilpotent orbits of simple Lie algebras. Then, in
Section \ref{sec:FLOWS}, we discuss the manifestation of Higgs branch
renormalization group flows between these SCFTs in terms of the underlying
bifiltered rings. Finally, we conclude and enumerate some questions and future directions in Section \ref{sec:DISC}. 

\section{Conjecture for the bifiltered ring of \texorpdfstring{\boldmath{$\mathcal{D}_p(SU(N), O)$}}{Dp(SU(N), O)}}\label{sec:conj}

As we have explained, in this paper we consider 4d $\mathcal{N}=2$ SCFTs where
the Higgs branch is a non-trivial algebraic variety. We focus on a particular
class of Argyres--Douglas theories known as $\mathcal{D}_p(G, O)$
\cite{Bonelli:2011aa,Cecotti:2012jx,Xie:2012hs,Cecotti:2013lda,Wang:2015mra,Wang:2018gvb}.
Here, we assume that $G = SU(N)$, and $p \geq 2$ is an integer with $\gcd(p, N)=1$. Here $O$ is a choice of nilpotent orbit of $\mathfrak{sl}(N, \mathbb{C})$, which labels the theory obtained by the corresponding nilpotent Higgsing of the `unHiggsed' $\mathcal{D}_p(G)$. 
See \cite{Couzens:2023kyf} for a review of this class of theories and the
conventions/notation that we follow. 
In this section, we conjecture the explicit bifiltered rings that underlie both the Higgs branch and the Macdonald index of these theories.  In particular, this conjecture makes manifest how the bifiltered rings vary along the Higgs branch RG flows.  We then test the conjecture extensively in the succeeding sections. 

The VOA associated to the Schur sector of the $\mathcal{D}_p(SU(N), O)$ theory is the W-algebra \cite{Xie:2016evu, Song:2017oew, Beem:2014rza} given by a Drinfeld--Sokolov reduction of the affine Kac--Moody algebra $\mathfrak{su}(N)_{k_{\text{2d}}}$ by the nilpotent orbit $O$: $W^{k_\text{2d}}(\mathfrak{su}(N), O)$, where $k_\text{2d}$ is fixed by the 4d flavor central charge. Following the discussion on VOAs in Section
\ref{sec:intro}, our conjecture is equivalently a conjecture for Zhu's $C_2$-algebra of each such simple W-algebra.

We start by defining a universal potential that is associated to the $\mathcal{D}_p(SU(N), O)$ family, with fixed $p$ and $N$. Consider the graded polynomial ring
\begin{equation}
    S_u = \mathbb{C}[u_2, \cdots, u_{N}] \,, \qquad \operatorname{deg}(u_i) = i \,.
\end{equation}
Let $P_u(u_2, \cdots, u_N)$ be a weighted homogeneous polynomial of degree $p +
1$ belonging to $S_u$. We want to find a $P_u$ which satisfies the following
properties. Consider the following ring of traceless, $N \times N$, matrices
over the complex numbers:
\begin{equation}
    S[M] \,, \qquad \operatorname{deg}(M_{k,\ell}) = 1 \,.
\end{equation}
This is the polynomial coordinate ring of $\mathfrak{sl}(N)$. Generally, we will write the matrix $M$ in terms of $N^2 - 1$ independent
coordinates that we refer to as $a_1, \cdots, a_{N^2 - 1}$. We can pull back
$P_u$ to the polynomial ring $S[M]$ as follows:
\begin{equation}
    P(M) = P_u\left( \, \frac{1}{2}\operatorname{tr}M^2 , \,\,\cdots,\,\, \frac{1}{N}\operatorname{tr}M^N \, \right) \,,
\end{equation}
and we define a quotient ring by taking the Jacobian ideal associated to $P$:
\begin{equation}
    R = S[M] \, / \, \left( \partial_{a_1} P ,\, \cdots,\, \partial_{a_{N^2-1}} P \right) \,.
\end{equation}
We assume that $P_u$ is such that $\operatorname{Spec}(R)_\text{red}$ is the
closure of the principal nilpotent orbit of $\mathfrak{sl}(N)$,\footnote{For $p < N$, the Higgs branch is not the full nilpotent cone, but the argument below goes through essentially unchanged. Instead, we require that $\operatorname{Spec}(R)_\text{red}$ is the Higgs branch of $\mathcal{D}_p(SU(N))$. Furthermore, for $p < N$, the Argyres--Douglas theory is not defined for arbitrary nilpotent orbit $O$, but only for nilpotent orbits satisfying certain properties. We return to this special case in Section \ref{sec:isos}.} and that there
exists an additional syzygy, compared to the generic case, of the first jet scheme of $R$ of $(q, p)$-bidegree $(p+2, 1)$. In particular, the multiplicity of the trivial $\mathfrak{su}(N)$ representation among the minimal second syzygies of bidegree $(p+2,1)$ is one:
\begin{equation}
    \beta_{2,(p+2, 1)}(J_1(R))\Big|_{\text{singlet}} = 1 \,.
\end{equation}
It is a non-trivial statement, which we defer to \cite{SYZYGY} for general $p$ and $N$, that this condition, together with the requirement of the correct reduced Higgs branch, produces a unique potential $P_u$, up to an overall rescaling which does not change the Jacobian ideal. We refer to this $P_u$ as the universal potential for $\mathcal{D}_p(SU(N), O)$.

Now that we have determined the potential $P_u$, which is independent of the
nilpotent orbit $O$, we are ready to incorporate the data of $O$. The Higgs
branch of $\mathcal{D}_p(SU(N))$, with $p > N$ and $\gcd(p, N) = 1$, is the
closure of the principal nilpotent orbit of $\mathfrak{sl}(N)$, also known as
the nilpotent cone: $\mathcal{N}$. Associated to each nilpotent orbit $O
\subseteq \mathcal{N}$ is a Slodowy slice, $\mathcal{S}_O$ for which the
intersection with the nilpotent cone, $\mathcal{N} \cap \mathcal{S}_O$,
captures the residual transverse Higgs branch geometry associated with
nilpotent Higgsing by $O$. More explicitly, choose a representative $e\in O$,
where $e$ is the explicit vacuum expectation value given to the moment map, and
complete it to an $\mathfrak{sl}(2)$ triple, $(e, h, f)$. The corresponding
Slodowy slice is
\begin{equation}
    \mathcal{S}_O = e + \operatorname{ker}(\operatorname{ad} f) \subseteq \mathfrak{g}_\mathbb{C} \,.
\end{equation}
Let $\{T_i\}$ form a basis of $\operatorname{ker}(\operatorname{ad} f)$ which
is homogeneous with respect to the adjoint action of $h$, and then a generic
element of the Slodowy slice can be written as
\begin{equation}
    M = e + \sum_i n_i T_i \,.
\end{equation}
We refer to the coefficients, $n_i$, as the affine coordinates on the Slodowy
slice. Since the basis consists of $h$-weight vectors, we can assign a
(half-)Kazhdan degree to each of the affine coordinates. If $[h, T_i] =
-\kappa_i T_i$, then we assign the degree as
\begin{equation}
    q_i = \operatorname{deg}(n_i) = 1 + \frac{\kappa_i}{2} \,.
\end{equation}
We denote the $\mathfrak{su}(N)$ Casimir invariants by
\begin{equation}
C_k(M)=\frac{1}{k}\operatorname{tr}M^k \,, \qquad k=2,\cdots, N \,,
\end{equation}
where $M$ is a traceless $N \times N$ matrix over $\mathbb{C}$. Restricting
these invariants to the Slodowy slice gives polynomials in the affine
coordinates, $n_i$:
\begin{equation}
    C_k|_{\mathcal{S}_O}(\bm n) = C_k\left(e+\sum_i n_iT_i\right) = \frac{1}{k}\operatorname{tr}\left(e+\sum_i n_iT_i\right)^k \,.
\end{equation}
These restricted Casimirs are homogeneous with respect to the half-Kazhdan grading:
\begin{equation}
    \operatorname{deg}\left( C_k|_{\mathcal{S}_O}(\bm{n})\right) = k \,.
\end{equation}

With this in hand, then, to determine the bifiltered ring for an arbitrary $O$,
we start by defining the polynomial ring:
\begin{equation}
    S_O = \mathbb{C}[n_1, \cdots, n_d, w_2, \cdots, w_N] \,, \qquad \operatorname{deg}(n_i) = (q_i, q_i) \,, \,\, \operatorname{deg}(w_i) = (i, i-1) \,,
\end{equation}
and define the polynomial $P_O$ belonging to $S_O$ via the pullback of the
universal potential:
\begin{equation}
    P_O(\bm{n}) = P_u(C_2|_{\mathcal{S}_O}(\bm{n}), \cdots, C_N|_{\mathcal{S}_O}(\bm{n})) \,.
\end{equation}
We can then define the relevant ideal of $S_O$ as
\begin{equation}\label{eqn:ringgen}
    I_O = \left( \partial_{n_1} P_O \,, \cdots \,, \partial_{n_d} P_O \,, w_2 -  C_2|_{\mathcal{S}_O}(\bm{n}) \,, \cdots \,, w_N -  C_N|_{\mathcal{S}_O}(\bm{n}) \right) \,.
\end{equation}
The generators involving $w_i$ are (in general) not homogeneous in
the $T$-grading, which leads to the bifiltration structure on the quotient.
Altogether, we propose that the bifiltered ring associated to the
Argyres--Douglas theory $\mathcal{D}_p(SU(N), O)$, with
$\gcd(p,N)=1$, is
\begin{equation}\label{eqn:conj}
    R_O = S_O \, / \, I_O \,.
\end{equation}
We conjecture that this ring is such that it realizes equations
\eqref{eqn:bigHIGGS} and \eqref{eqn:bigMAC}.

If we forget about the $T$-grading, then the linear equations for the $w_i$ become homogeneous in the only remaining grading, the $q$-grading. Then, we can solve the linear equations in $I_O$ to write a ring $R_{S,O}$, which is equivalent to $R_O$ as a $q$-graded ring.
We therefore also conjecture that this ring is Zhu's $C_2$-algebra for the simple $W$-algebras: $W^{k_\text{2d}}(\mathfrak{su}(N), O)$.

\section{Bifiltered rings for non-point Higgs branches}\label{sec:RINGS}

We now determine the bifiltered rings explicitly for several families of Argyres--Douglas theories with non-trivial Higgs branches. Generalizing from examples with point-like Higgs branches studied in \cite{Kang:2025zub,Kang:2026nge}, the essential point to note is that the reduced Higgs branch geometry $\operatorname{Spec}(R)_{\mathrm{red}}=\mathcal M_H$ must be imposed as part of the reconstruction. In the point-like cases, this constraint imposed that the bifiltered rings be complete intersection rings.

Our reconstruction proceeds as follows. From the Schur index, together with the flavor symmetry, we identify the generators and their $q$- and flavor gradings, as well as the degrees and representation content of the expected relations. We then construct the most general symmetry-compatible potential $P$ whose Jacobian ideal has the required reduced spectrum. This leaves a finite-dimensional family of candidate schemes with the same Higgs branch. We select among them by imposing syzygy maximization: the first jet scheme must acquire the distinguished minimal second syzygy $\beta_{2,(\ell,1)}\!\left(J_1(R_S)\right)=1$ in the appropriate flavor-singlet sector. In the examples below, this condition fixes the otherwise undetermined coefficients of the potential. We then test the resulting ring by computing the Hilbert series of its arc space and comparing it with the Macdonald index whenever the latter is known, or with its Schur and Hall--Littlewood limits otherwise.

In fact, we have determined closed-form expressions for the syzygy-maximized potentials associated to the $(A_1, A_\text{odd})$ and $\mathcal{D}_p(SU(N))$ families of theories \cite{SYZYGY}.\footnote{And thus, following the conjecture in Section \ref{sec:conj}, for all $\mathcal{D}_p(SU(N), O)$.} In this section, we consider only theories where we have convincingly and explicitly demonstrated that the ideal associated to this syzygy-maximized potential, as in equation \eqref{eqn:ringgen}, reproduces either the Macdonald index, or known limits of the Macdonald index, via the Hilbert series of the arc space. 
This is an explicit computation for fixed numerical parameters; we still need to perform it with parametric dependence across the general families for which we construct the syzygy-maximized potentials in \cite{SYZYGY}. Thus, we do not stress herein the calculational process by which we determine the syzygy-maximized potentials; we defer this to \cite{SYZYGY}.

We begin with theories whose Higgs branches are $A$-type orbifold singularities $\mathbb C^2/\mathbb Z_n$. These include the $(A_1,A_{2n-1})$ theories in Section \ref{sec:A1A2nm1} and the subregularly Higgsed theories $D_p(SU(n),[n-1,1])$ in Section \ref{sec:DpSUnRG}. In Section \ref{sec:DpSUN}, we turn to the unHiggsed $D_p(SU(N))$ theories, whose Higgs branches are closures of nilpotent orbits rather than orbifold singularities. These examples demonstrate how the reduced Higgs branch geometry and the protected operator spectrum work together: the former constrains the underlying classical variety, while syzygy maximization selects the additional scheme-theoretic structure required to reproduce the index.

\subsection{The \texorpdfstring{{$(A_1, A_{2n-1})$}}{(A1, A2n-1)} Argyres--Douglas SCFTs}\label{sec:A1A2nm1}

We begin by deriving the bifiltered rings for the $(A_1, A_{2n-1})$ Argyres--Douglas theories via syzygy maximization. The starting point is the Schur index of the $(A_1, A_{2n-1})$ theory, which was given in \cite{Buican:2015ina, Song:2017oew} as 
\begin{align}
    I(q; z) = \textrm{PE} \left[ \frac{q + q^2 +  q^{\frac{n}{2}}(z^2+\frac{1}{z^2}) - q^{\frac{n}{2}+2} ( z^2 + \frac{1}{z^2}) -q^n -q^{n+1}}{(1-q) \left(1-q^{n+1}\right)} \right] \,.
\end{align}
The theory possesses a $\mathfrak{u}(1)$ symmetry, which we capture in the index via the fugacity $z$. From the index, we observe four generators of $(q, z)$-bidegrees:
\begin{equation}\label{eqn:ripto}
    (n/2, 2) \,, \quad (n/2, -2) \,, \quad (1,0) \,,\quad  (2,0) \,.
\end{equation}
We can also observe that there are relations of $(q, z)$-bidegrees
\begin{equation}
    (n,0) \,, \quad (n+1,0) \,, \quad (n/2 + 2, 2) \,, \quad (n/2 + 2, -2) \,.
\end{equation}
Therefore, if we ask for the relations to be reproduced from a Jacobian ideal of a potential, $P$, then we would expect $P$ to have $(q, z)$-bidegree $(n+2, 0)$. Further requiring that the reduced spectrum of the quotient ring arising from the Jacobian ideal is $\mathbb{C}^2/\mathbb{Z}_n$, we find that $P$ is forced to take the following form:
\begin{equation}\label{eqn:A1A2nm1POLY}
    P_{n+2}(x, y, z, t) = (xy+z^n)t + \sum_{i=1}^{\lfloor {n/2} \rfloor} \alpha_i z^{n-2i} t^{i+1} \,,
\end{equation}
which belongs to the polynomial ring $\mathbb{C}[x, y, z, t]$,
where the coordinates have bidegrees as in equation \eqref{eqn:ripto}. This defines a family of Jacobian ideals parametrized by the coefficients $\alpha_i$:
\begin{equation}\label{eqn:bianca}
    R_S = \mathbb{C}[x, y, z, t] \, / \,  \left( \partial_x P_{n+2}, \partial_y P_{n+2}, \partial_z P_{n+2}, \partial_t P_{n+2} \right) \,.
\end{equation}
Our task is to determine these coefficients by maximizing the syzygies and then verify that the quotient ring by the resulting ideal reproduces the Schur index via the proposal in equation \eqref{eqn:bigSCHUR}.
The closed-form expression for the Schur index can be expanded to indicate that a $P_{n+2}$ which leads to an $R_S$ realizing equation \eqref{eqn:bigSCHUR} must satisfy
\begin{equation}
    \beta_{2,(n+3,1)}(J_1(R_S)) \Big|_{\text{singlet}} = 1 \,.
\end{equation}
That is, the coordinate ring of the first jet scheme must contain an independent syzygy of $(q,p)$-bidegree $(n+3,1)$ in the flavor-singlet sector, equivalently at $z$-degree zero. A systematic determination of the $\alpha_i$ satisfying this syzygy condition for general values of $n$ is given in \cite{SYZYGY}, and, in each case, the syzygy condition uniquely determines $R_S$.

In fact, the Macdonald indices for the $(A_1, A_{2n-1})$ theories are also known \cite{Buican:2015tda}, and thus we can simply input the necessary $T$-grading. Altogether, we consider the ring
\begin{equation}
  \begin{gathered}
    R = \mathbb{C}[x, y, z, t] \, / \, \left( \partial_x P, \partial_y P, \partial_z P, \partial_t P \right) ,\\[0.5em]
    \operatorname{deg}(x) = \left( \frac{n}{2}, \frac{n}{2}, 2 \right) , \,\, \operatorname{deg}(y) = \left( \frac{n}{2}, \frac{n}{2}, -2 \right) , \,\, \operatorname{deg}(z) = (1, 1, 0) , \,\, \operatorname{deg}(t) = (2, 1, 0) \,,
  \end{gathered}
\end{equation}
where we have written the $(q, T, z)$-tridegrees in that order, and where $P$ is of the form in equation \eqref{eqn:A1A2nm1POLY}, with all $\alpha_i$ fixed by syzygy maximization. Then, for each of the $(A_1, A_{2n-1})$ in Table \ref{tab:A1Aodd}, we find that the Hilbert series of the arc space of $R$ reproduces the Macdonald index as per equation \eqref{eqn:bigMAC}.\footnote{The Macdonald index computations are performed to approximately order $q^{n + 6}$.}

While we have here provided sufficient information for the reader to reconstruct the Macdonald indices for each of these theories from the Hilbert series of the arc space (to arbitrary order, given sufficient computational resources), we present here just one explicit example. For the $(A_1, A_5)$ SCFT, the Macdonald index, as computed via equation \eqref{eqn:bigMAC}, is:

\vspace{-12pt}
\scriptsize
\begin{align*}
I_M{}&{}^{(A_1, A_5)}=1+qT+2q^{3/2}T^{3/2}
+q^2(2T+T^2)+q^{5/2}(2T^{3/2}+2T^{5/2})+q^3(2T+2T^2+3T^3)+q^{7/2}(2T^{3/2}+4T^{5/2}+2T^{7/2})\\
&+q^4(2T+5T^2+4T^3+3T^4)+q^{9/2}(2T^{3/2}+8T^{5/2}+4T^{7/2}+4T^{9/2})+q^5(2T+6T^2+9T^3+6T^4+3T^5)\\
&+q^{11/2}(2T^{3/2}+12T^{5/2}+10T^{7/2}+6T^{9/2}+4T^{11/2})+q^6(2T+9T^2+14T^3+13T^4+6T^5+5T^6)\\
&+q^{13/2}(2T^{3/2}+16T^{5/2}+18T^{7/2}+14T^{9/2}+8T^{11/2}+4T^{13/2})+q^7(2T+10T^2+22T^3+24T^4+15T^5+8T^6+5T^7)\\
&+q^{15/2}(2T^{3/2}+20T^{5/2}+32T^{7/2}+26T^{9/2}+18T^{11/2}+8T^{13/2}+6T^{15/2})\\
&+q^8(2T+13T^2+30T^3+42T^4+28T^5+19T^6+10T^7+5T^8)\\
&+q^{17/2}(2T^{3/2}+24T^{5/2}+48T^{7/2}+46T^{9/2}+34T^{11/2}+20T^{13/2}+10T^{15/2}+6T^{17/2})\\
&+q^9(2T+14T^2+42T^3+64T^4+54T^5+36T^6+23T^7+10T^8+7T^9)\\
&+q^{19/2}(2T^{3/2}+28T^{5/2}+70T^{7/2}+74T^{9/2}+64T^{11/2}+38T^{13/2}+24T^{15/2}+12T^{17/2}+6T^{19/2})\\
&+q^{10}(2T+17T^2+52T^3+97T^4+90T^5+68T^6+44T^7+25T^8+12T^9+7T^{10})\\
&+q^{21/2}(2T^{3/2}+32T^{5/2}+94T^{7/2}+120T^{9/2}+106T^{11/2}+74T^{13/2}+46T^{15/2}+28T^{17/2}+12T^{19/2}+8T^{21/2})\\
&+q^{11}(2T+18T^2+66T^3+136T^4+149T^5+114T^6+84T^7+48T^8+29T^9+14T^{10}+7T^{11})\\
&+q^{23/2}(2T^{3/2}+36T^{5/2}+124T^{7/2}+176T^{9/2}+176T^{11/2}+128T^{13/2}+88T^{15/2}+54T^{17/2}+30T^{19/2}+14T^{21/2}+8T^{23/2})\\
&+q^{12}(2T+21T^2+80T^3+188T^4+226T^5+195T^6+144T^7+94T^8+56T^9+33T^{10}+14T^{11}+9T^{12})
 +O(q^{25/2}) \,.
%
\end{align*}
\normalsize

\begin{table}[H]
    \centering
    \begin{threeparttable}
    $\begin{array}{c @{\hskip 0.3in} l}
    \toprule
    (A_1, A_{2n-1}) & P_{n+2} (x, y, z, t)  \\
    \midrule
    (A_1, A_3) & P_4 = (xy+z^2)t + t^2 \\
    (A_1, A_5) & P_5 = (xy+z^3)t + z t^2 \\
    (A_1, A_7) & P_6 = (xy+z^4)t + z^2 t^2 + \frac{1}{18} t^3 \\
    (A_1, A_9) & P_7 = (xy+z^5)t + z^3 t^2 + \frac{1}{10} z t^3 \\
    (A_1, A_{11}) & P_8 = (xy + z^6) t + z^4 t^2 + \frac{2}{15} z^2 t^3 + \frac{1}{675} t^4    \\
    (A_1, A_{13}) & P_9 = (xy+z^7) t + z^5 t^2 + \frac{10}{63} z^3 t^3 + \frac{5}{1323} z t^4 \\
    (A_1, A_{15}) & P_{10} = (x y+z^8)t + z^6 t^2  + \frac{5}{28} z^4 t^3  + \frac{5}{784} z^2 t^4 + \frac{1}{43904} t^5 \\
    (A_1, A_{17}) & P_{11} = (xy + z^9)t + z^7 t^2 + \frac{7}{36} z^5 t^3 + \frac{35}{3888} z^3 t^4 + \frac{7}{93312} z t^5 \\
    (A_1, A_{19}) & P_{12} = (x y + z^{10})t + z^8 t^2  + \frac{28}{135} z^6 t^3 + \frac{14}{1215} z^4 t^4 + \frac{14}{91125} z^2 t^5 + \frac{14}{61509375} t^6 \\\bottomrule 
    \end{array}$
    \end{threeparttable}
    \caption{The syzygy-maximized potentials underlying the Jacobian ideal for the bifiltered affine schemes that encode both the Macdonald indices and the $\mathbb{C}^2/\mathbb{Z}_n$ Higgs branches of the $(A_1, A_{2n-1})$ Argyres--Douglas theories. We have reparametrized the variables to minimize the number of non-unit coefficients.}
    \label{tab:A1Aodd}
\end{table}

\subsection{More \texorpdfstring{$\mathbb{C}^2/\mathbb{Z}_n$}{C2/Zn}: \texorpdfstring{$\mathcal{D}_p(SU(n), [n-1, 1])$}{Dp(SU(n), [n-1,1])} Argyres--Douglas SCFTs} \label{sec:DpSUnRG}

We next consider a further set of SCFTs where the Higgs branch is an A-type orbifold singularity. These are the $D_p(SU(n), [n-1,1])$ theories. For $\gcd(p, n) = 1$ and $p > n$, the corresponding VOA is the W-algebra $W_{k_{2d}} (\mathfrak{su}_n, [n-1,1])$ with $k_{2d} =-n+n/p$. For $p = n+1$, these theories are the same as the $(A_1, A_{2n-1})$ SCFTs discussed in Section \ref{sec:A1A2nm1}.

We first consider the case of $n = 3$. The Schur index for this class of theories has a closed-form in terms of a plethystic exponential \cite{Song:2017oew}:
\begin{align}\label{eqn:sparx}
    I_S (q, z) = \textrm{PE} \left[\frac{q+q^{\frac32}(z+z^{-1})+q^2 - q^{p-1} - q^{p-\frac12}(z+z^{-1}) - q^p}{(1-q)(1-q^p)} \right] \,. 
\end{align}
The structure of the Schur index indicates that we should consider a bigraded ring (where the second grading is for a $\mathfrak{u}(1)$ flavor symmetry) of the form:
\begin{equation}\label{eqn:elora}
    \begin{gathered}
        R_S = \mathbb{C}[x,y,z,t] \, / \, I_{P_{p+1}} \\[0.5em]
        \operatorname{deg}(x) = \left( \frac{3}{2}, 1 \right) , \,\, \operatorname{deg}(y) = \left( \frac{3}{2}, -1 \right) , \,\, \operatorname{deg}(z) = (1, 0) , \,\, \operatorname{deg}(t) = (2, 0) \,,
    \end{gathered}
\end{equation}
where $I_{P_{p+1}}$ is the Jacobian ideal associated to a homogeneous polynomial of $q$-degree $p + 1$. The condition that $\operatorname{Spec}(R_S)_\text{red}$ is isomorphic to $\mathbb{C}^2 \, / \, \mathbb{Z}_3$ enforces that the potential $P_{p+1}$ cannot have the form of a generic polynomial in these four coordinates, but must be written in terms of an auxiliary coordinate
\begin{equation}\label{eqn:boots}
     w = xy+z^3 + zt \,.
\end{equation}
The ansatz for the potential is then
\begin{align}
    P_{p+1} (w, t) = \sum_{3a+2b=p+1} \alpha_{a,b}w^a t^b  \,,
\end{align}
where the $\alpha_{a,b}$ are yet-to-be-determined coefficients. These coefficients are fixed by the syzygy maximization requirement that
\begin{equation}
    \beta_{2,(p+2,1)}(J_1(R_S)) \Big|_\text{singlet} = 1 \,,
\end{equation}
and the $R_S$ obtained via such impositions is unique. For explicit values of $p$, we find the corresponding potentials as written in Table \ref{tab:DpSU3RG} by performing this maximization. Indeed, the resulting rings $R_S$ reproduce the Schur index of the theories, as in equation \eqref{eqn:sparx}, via the Hilbert series of the arc space in equation \eqref{eqn:bigSCHUR}.

\vspace{0.6cm}
\begin{table}[H]
    \centering
    \begin{threeparttable}
    $\begin{array}{c @{\hskip 0.3in} l}
    \toprule
    \mathcal{D}_p(SU(3), [2,1]) & P_{p+1}(w,t)\\
    \midrule
    p=4 & P_5 (w, t) = w t \\
    p=5 & P_6 (w, t) = w^2 - \frac{1}{9} t^3 \\
    p=7 & P_8 (w, t) = w^2 t - \frac{1}{18}t^4 \\
    p=8 & P_9 (w, t) = w^3 - \frac{1}{3} w t^3 \\
    p=10 & P_{11}(w, t) = w^3 t - \frac{1}{6} w t^4 \\
    p=11 & P_{12}(w, t) = w^4 - \frac{2}{3} w^2 t^3 + \frac{2}{135} t^6 \\\bottomrule
    \end{array}$
    \end{threeparttable}
    \caption{The potentials capturing the bifiltered rings, as determined by syzygy maximization, for the $\mathcal{D}_p(SU(3), [2,1])$ Argyres--Douglas theories.}
    \label{tab:DpSU3RG}
\end{table}
\vspace{1cm}

For $p > 4$, the Macdonald index of $\mathcal{D}_p(SU(3), [2,1])$ is not generally known. For $p = 5$, the theory is dual to the $(A_1, E_7)$ \cite{Song:2017oew},\footnote{There are a couple of typos in \cite{Song:2017oew} when identifying the $(A_1, E_7)$ theory.} and the Macdonald index has recently been computed in \cite{Kim:2025klh}. 
There is a natural extension of the ring $R_S$ defined in equation \eqref{eqn:elora}, to a bifiltered ring incorporating a $T$-grading.\footnote{Actually it is trifiltered when remembering the $z$-grading of the additional $\mathfrak{u}(1)$ flavor symmetry.} This is
\begin{equation}
    \begin{gathered}
        R = \mathbb{C}[x,y,z,t,w] \, \big/ \, \big( I_{P_{p+1}} + (w - xy - z^3 - zt) \big) \,,
    \end{gathered}
\end{equation}
where
\begin{equation}
    \begin{gathered}
        \operatorname{deg}(x) = \left( \frac{3}{2},  \frac{3}{2}, 1 \right) , \,\, \operatorname{deg}(y) = \left( \frac{3}{2},  \frac{3}{2}, -1 \right) , \\[0.3em] \operatorname{deg}(z) = (1, 1, 0) , \,\,
        \operatorname{deg}(t) = (2, 1, 0) , \,\,\operatorname{deg}(w) = (3, 2, 0) \,,
    \end{gathered}
\end{equation}
and where $P_{p+1}$ is the same potential that was derived from the procedure of syzygy maximization, now written directly in terms of the new coordinate $w$, and $t$. We have also promoted the auxiliary relation in equation \eqref{eqn:boots} to a generator of the ideal. 
For $p = 4$ and $p = 5$, this reproduces the Macdonald indices of \cite{Buican:2015tda,Kim:2025klh} via equation \eqref{eqn:bigMAC}. For the $(A_1, E_7)$ SCFT, the Hilbert series of the arc space gives the following Macdonald index:

\vspace{-12pt}\scriptsize
\begin{align*}
\begin{aligned}
I_M{}&{}^{D_5(SU(3),[2,1])=(A_1,E_7)}=1+qT+2q^{3/2}T^{3/2} +q^2(2T+T^2)+q^{5/2}(2T^{3/2}+2T^{5/2})\\
&+q^3(2T+3T^2+3T^3)+q^{7/2}(2T^{3/2}+6T^{5/2}+2T^{7/2}) +q^4(2T+6T^2+5T^3+3T^4)+q^{9/2}(2T^{3/2}+10T^{5/2}+6T^{7/2}+4T^{9/2})\\
&+q^5(2T+7T^2+12T^3+9T^4+3T^5)+q^{11/2}(2T^{3/2}+14T^{5/2}+16T^{7/2}+8T^{9/2}+4T^{11/2})\\
&+q^6(2T+10T^2+20T^3+20T^4+9T^5+5T^6) +q^{13/2}(2T^{3/2}+18T^{5/2}+30T^{7/2}+20T^{9/2}+12T^{11/2}+4T^{13/2})\\
&+q^7(2T+11T^2+30T^3+38T^4+24T^5+11T^6+5T^7)\\
&+q^{15/2}(2T^{3/2}+22T^{5/2}+50T^{7/2}+42T^{9/2}+28T^{11/2}+12T^{13/2}+6T^{15/2})\\
&+q^8(2T+14T^2+40T^3+67T^4+50T^5+28T^6+15T^7+5T^8)\\
&+q^{17/2}(2T^{3/2}+26T^{5/2}+72T^{7/2}+80T^{9/2}+58T^{11/2}+32T^{13/2}+14T^{15/2}+6T^{17/2})\\
&+q^9(2T+15T^2+54T^3+103T^4+101T^5+60T^6+36T^7+15T^8+7T^9)\\
&+q^{19/2}(2T^{3/2}+30T^{5/2}+100T^{7/2}+136T^{9/2}+114T^{11/2}+68T^{13/2}+36T^{15/2}+18T^{17/2}+6T^{19/2})\\
&+q^{10}(2T+18T^2+66T^3+153T^4+175T^5+123T^6+76T^7+40T^8+17T^9+7T^{10})\\
&+q^{21/2}(2T^{3/2}+34T^{5/2}+130T^{7/2}+218T^{9/2}+202T^{11/2}+142T^{13/2}+78T^{15/2}+44T^{17/2}+18T^{19/2}+8T^{21/2})\\
&+q^{11}(2T+19T^2+82T^3+211T^4+290T^5+227T^6+155T^7+86T^8+44T^9+21T^{10}+7T^{11})\\
&+q^{23/2}(2T^{3/2}+38T^{5/2}+166T^{7/2}+320T^{9/2}+346T^{11/2}+264T^{13/2}+162T^{15/2}+94T^{17/2}+48T^{19/2}+20T^{21/2}+8T^{23/2})\\
&+q^{12}(2T+22T^2+98T^3+284T^4+444T^5+406T^6+287T^7+181T^8+96T^9+52T^{10}+21T^{11}+9T^{12})
 +O(q^{25/2})\,,
\end{aligned}
\end{align*}
\normalsize
where we emphasize that this goes far beyond the order computed in \cite{Kim:2025klh}. For $p > 5$, where the Macdonald index is unknown, we verify that the Hall--Littlewood limit of the Hilbert series of the arc space of $R$ reproduces the known Hall--Littlewood index \cite{Song:2017oew}, which matches with the Higgs branch Hilbert series.

It is instructive to interpret these results from the VOA perspective. The VOA associated to $\mathcal D_p(SU(3),[2,1])$ is the simple Bershadsky--Polyakov algebra at level \cite{MR1116410,Polyakov:1989dm,Feigin:2004wb}
\begin{align}
    k_{2d}=-3+\frac{3}{p} \,,
\end{align}
with central charge
\begin{align}
    c_{2d}=-\frac{(2k_{2d}+3)(3k_{2d}+1)}{k_{2d}+3} \,.
\end{align}
Writing $k_{2d}+3=u/v$, with $u$ and $v$ coprime, an admissible level is called nondegenerate when $u,v\geq 3$. The theories considered here correspond to $(u,v)=(3,p)$ and therefore, for $p>3$, lie precisely in the nondegenerate-admissible regime. Unlike the exceptional admissible series with $v=2$, the corresponding simple Bershadsky--Polyakov algebras are neither lisse nor rational \cite{Fehily:2021sdl,Fehily:2020bif,MR4801934}. Their non-lisse character is consistent with the positive-dimensional associated variety $\mathbb C^2/\mathbb Z_3$ predicted by the Higgs branch of the four-dimensional theory.

Although the representation theory and modular properties of Bershadsky--Polyakov algebras at nondegenerate-admissible levels have been studied extensively, the vacuum null ideal, and hence the full scheme-theoretic structure of the $C_2$-algebra, is generally difficult to determine explicitly. Under the proposed identification of $R_S$ with the $C_2$-algebra of the associated VOA, the potentials in Table \ref{tab:DpSU3RG} provide explicit presentations of this structure for the family $\mathrm{BP}(3,p)$. 
Equivalently, the syzygy-maximized relations predict the commutative images of the vacuum null relations, while their derivative descendants generate the relations needed to reproduce the Schur index. Thus, the protected data of the four-dimensional SCFT supplies explicit scheme-theoretic information about these non-rational Bershadsky--Polyakov models that is not readily accessible from their existing representation-theoretic descriptions.

Having established the construction for $\mathbb C^2/\mathbb Z_3$, we now ask whether it extends to the higher orbifold singularities $\mathbb C^2/\mathbb Z_n$ with $n>3$.
The Schur index for this class of theories is given as
\begin{align}
\begin{split}
    I_S(q, z) &= \text{PE} \left[ \frac{q+q^2+\cdots + q^{n-1} + q^{\frac n2}(z+\frac{1}{z}) - q^{p-n+2} - \cdots - q^p - q^{p+1-\frac{n}{2}} (z+\frac{1}{z})}{(1-q)(1-q^p)} \right] \,.
\end{split}
\end{align}
We see that for $p=n+1$, we get the Schur index of $(A_1, A_{2n-1})$ theory. 
If we assume the ring has the structure of a Jacobian ideal, and the same pattern as the previous example, we expect the defining potential to be a polynomial
\begin{align}
    P_{p+1}(x, y, z, t, w_3, w_4, \cdots w_{n-1}) \ , 
\end{align}
of degree $p+1$, and satisfying the constraint
\begin{align}
    \beta_{2, (p+2, 1)}(J_1(R_S)) \Big|_{\text{singlet}} = 1 \,. 
\end{align}
Notice that we have introduced higher-spin coordinates $w_3, w_4, \cdots w_{n-1}$ of degrees $3, 4, \cdots, n-1$, in addition to the usual $x, y, z, t$ of degrees $n/2, n/2, 1, 2$, to match the structure apparent in the Schur index. The Jacobian ideal of this potential contains generators of degrees $p, p-1, \cdots, p-n+2$ and $p+1-n/2$, matching with the expectations from the Schur index. 

For example, consider $n = 4$ and $p=7$, then, using the syzygy maximization condition, we find that all free coefficients in the potential $P_{p+1}$ are solved for, and we obtain a unique quotient ring, $R_S$. The potential we find is:
\begin{align} \label{eq:D7su4Higgs}
    P_8 = (xy + z^4 + z^2t + zw)^2 + \frac{1}{4}(xy + z^4 + z^2t + zw) t^2 - \frac{1}{4} t w^2 + \frac{5}{192} t^4 \,.
\end{align}
It is also straightforward to see that the spectrum of the associated quotient ring reduced to a variety reproduces $\mathbb{C}^2 \, / \, \mathbb{Z}_4$, the expected Higgs branch. We can uplift the graded ring $R_S$ obtained in such a way to a bifiltered ring $R$ by introducing new coordinates with linear, but inhomogeneous under the $T$-grading, relations, exactly as we did for $n = 3$. The Hilbert series of the arc space then produces a putative Macdonald index:

\vspace{-12pt}\scriptsize
\begin{align*}
\begin{aligned}
I_M{}&{}^{\mathcal{D}_7(SU(4),[3,1])}=1+qT +q^2(2T+3T^2) +q^3(2T+5T^2+3T^3) +q^4(2T+8T^2+10T^3+5T^4)\\
&+q^5(2T+9T^2+19T^3+14T^4+5T^5) +q^6(2T+12T^2+30T^3+36T^4+18T^5+7T^6)\\
&+q^7(2T+13T^2+41T^3+63T^4+48T^5+22T^6+7T^7) +q^8(2T+16T^2+54T^3+105T^4+107T^5+62T^6+26T^7+9T^8)\\
&+q^9(2T+17T^2+69T^3+149T^4+196T^5+141T^6+72T^7+30T^8+9T^9)\\
&+q^{10}(2T+20T^2+84T^3+211T^4+322T^5+295T^6+175T^7+86T^8+34T^9+11T^{10})\\
&+q^{11}(2T+21T^2+101T^3+277T^4+489T^5+533T^6+377T^7+205T^8+96T^9+38T^{10}+11T^{11})\\
&+q^{12}(2T+24T^2+120T^3+362T^4+705T^5+905T^6+742T^7+457T^8+237T^9+110T^{10}+42T^{11}+13T^{12})
 +O(q^{13}) \,.
\end{aligned}
\end{align*}
\normalsize
The Macdonald index of this theory has not been determined before, however we can verify that the $q^mT^m$ terms in this expression reproduce the known Hall--Littlewood index of the SCFT.

\subsection{The \texorpdfstring{{$\mathcal{D}_p(SU(N))$}}{Dp(SUN)} Argyres--Douglas SCFTs}\label{sec:DpSUN}

Finally, we turn to a family of examples where the Higgs branch is both nontrivial and not an orbifold singularity. We consider the Argyres--Douglas theories $\mathcal{D}_p(SU(N))$ with $\gcd(p,N) = 1$. It is particularly imperative to determine the potentials underlying the bifiltered rings for this class of theories; we can then use the nilpotent Higgsing conjecture in Section \ref{sec:conj} to determine the bifiltered rings for all $\mathcal{D}_p(SU(N), O)$, with any allowed nilpotent orbit $O$. This then reproduces all the potentials from Sections \ref{sec:A1A2nm1} and \ref{sec:DpSUnRG}. A general determination of the potentials for generic such $p$ and $N$ is given in \cite{SYZYGY}; here we restrict to the cases $N=3,4,5$ where we explicitly match the Hilbert series of their arc spaces to the known expressions or limits of the Macdonald index.
\begin{table}[H]
    \centering
    \begin{threeparttable}
    $\begin{array}{c @{\hskip 0.2in} c @{\hskip 0.3in} l}
    \toprule
    \multicolumn{2}{@{\hskip 0.1in} c @{\hskip 0.3in}}{\mathcal{D}_p(SU(N))} & P_{p+1}(\vec{a})\\
    \midrule
    \multirow{7}{*}{$N=3$} & p = 2 & P_3(\vec{a}) = C_3 \\
    & p = 4 & P_5(\vec{a}) = C_2 C_3 \\
    & p = 5 & P_6(\vec{a}) = C_3^2 + \frac{1}{9} C_2^3 \\
    & p = 7 & P_8(\vec{a}) = C_3^2 C_2 + \frac{1}{18} C_2^4  \\
    & p = 8 & P_9(\vec{a}) = C_3^3 + \frac{1}{3} C_3 C_2^3  \\
    & p = 10 & P_{11}(\vec{a}) = C_3^3 C_2 + \frac{1}{6} C_3 C_2^4   \\
    & p = 11 & P_{12}(\vec{a}) = C_3^4 + \frac{2}{3} C_3^2 C_2^3 + \frac{2}{135} C_2^6 \\\midrule
    \multirow{3}{*}{$N=4$} & p = 3, & P_4(\vec{a})  = C_4 - \frac{3}{8} C_2^2 \\
    & p = 5 & P_6(\vec{a})  = C_4 C_2 + \frac{1}{2} C_3^2 - \frac{3}{8} C_2^3 \\
    & p = 7 & P_8(\vec{a})  = C_4^2 - \frac{3}{4} C_4 C_2^2 + \frac{1}{4} C_3^2 C_2 + \frac{29}{192} C_2^4 \\\midrule
    \multirow{4}{*}{$N=5$} & p = 2 & P_3(\vec{a})  = C_3 \\
    & p = 3 & P_4(\vec{a})  = C_4 - \frac{3}{10} C_2^2 \\
    & p = 4 & P_5(\vec{a})  = C_5 - \frac{4}{5} C_3 C_2 \\
    & p = 6 & P_7(\vec{a})  = C_5 C_2 + C_4 C_3 - \frac{11}{10} C_3 C_2^2 \\\bottomrule
    \end{array}$
    \end{threeparttable}
    \caption{For $\mathcal{D}_p(SU(N))$, with various values of $p$ and $N$, we determined the bifiltered affine schemes that underlie the Macdonald index via equation \eqref{eqn:bigMAC}. They are specified by a potential $P$, via the mechanism in equation \eqref{eqn:ringgen}. The potentials, defined over the polynomial ring in the coordinates $a_1, \cdots, a_{N^2-1}$ (collectively $\vec{a}$),  given in this table are determined from syzygy maximization. See \cite{SYZYGY} for more details.}
    \label{tab:DpSUN}
\end{table}

\paragraph{\boldmath{$\mathcal{D}_p(SU(3))$}:} We find that the Schur indices for the $\mathcal{D}_p (SU(3))$, for low values of $p$, are reproduced by the potentials in Table \ref{tab:DpSUN}, where $C_2(\vec{a}) = \mathrm{Tr} M^2/2 $ and $C_3(\vec{a}) = \mathrm{Tr} M^3/3$ are quadratic and cubic Casimirs of $\mathfrak{su}(3)$, respectively. 
We find that each $\mathcal{D}_p(SU(3))$ corresponds to the degree-$(p+1)$ potential, whose coefficients are fixed by demanding a syzygy of bidegree $(p+2, 1)$ in the $\mathfrak{su}(3)$ singlet sector. 
In order to obtain the Macdonald index, we introduce new generators $t, w$ with the additional relations
\begin{align}
    t = C_2(\vec{a}) \ , \quad  w = C_3(\vec{a}) \,,
\end{align} 
with $T$-grading $T(t)=1, T(w)=2$. These defining relations for $t, w$ are not homogeneous with respect to the $T$-grading and also not included in the Jacobian of the potential $P$. However, altogether, this gives rise to the desired bifiltered ring, given as
\begin{align}
  \mathbb{C}[a_1, \cdots a_8, t, w]/(\partial_{a_1} P, \cdots, \partial_{a_8} P, t-C_2(\vec{a}), w - C_3(\vec{a}) )  \,.
\end{align}
Armed with this, we construct the arc space and its associated graded vector space, and then compute the Hilbert series to obtain the Macdonald index. We give here the Hilbert series of the arc space for $p = 4$:

\vspace{-12pt}
\scriptsize
\begin{equation*}
\begin{aligned}
I_M{}&{}^{\mathcal{D}_4(SU(3))}=1+q(8T) +q^2(9T+35T^2) +q^3(9T+72T^2+111T^3) ~+q^4(9T+117T^2+306T^3+286T^4) \\
&~+q^5(9T+153T^2+665T^3+936T^4+637T^5) ~+q^6(9T+198T^2+1118T^3+2422T^4+2331T^5+1274T^6)\\
&~+q^7(9T+234T^2+1718T^3+4904T^4+6741T^5+5040T^6+2346T^7)\\
&~+q^8(9T+279T^2+2411T^3+8803T^4+15450T^5+15722T^6+9828T^7+4047T^8)\\
&~+q^9(9T+315T^2+3260T^3+14203T^4+31330T^5+39183T^6+32374T^7+17712T^8+6622T^9)\\
&~+q^{10}(9T+360T^2+4193T^3+21625T^4+56781T^5+86659T^6+85614T^7+60777T^8+29997T^9+10373T^{10})
 +O(q^{11}) \,.
\end{aligned}
\end{equation*}
\normalsize

\paragraph{\boldmath{$\mathcal{D}_p(SU(4))$}:}
A similar analysis can be performed for the $\mathcal{D}_p (SU(4))$ family of theories. The known Schur indices are reproduced by potentials in Table \ref{tab:DpSUN}. As usual, $C_k = \mathrm{Tr}(M^k)/k$ is the $k$th Casimir invariant, which is a function of the fifteen generators of the adjoint representation of $\mathfrak{su}(4)$.
In order to obtain the Macdonald index, we introduce new generators $t, w_3, w_4$, together with the additional relations
\begin{align}
    t = C_2(\vec{a}) \ , \quad  w_3 = C_3(\vec{a}) \ , \quad w_4 = C_4 (\vec{a}) \,.
\end{align} 
The $T$-grading of these new coordinates is specified by $T(t)=1, T(w_3)=2, T(w_4)=3$. The bifiltered ring is then
\begin{align}
  \mathbb{C}[a_1, \cdots a_{15}, t, w_3, w_4]/(\partial_{a_1} P, \cdots, \partial_{a_{15}} P, t-C_2(\vec{a}), w_3 - C_3(\vec{a}), w_4 - C_4(\vec{a}) )  \,.
\end{align}
We have explicitly computed the Hilbert series of the arc space of this bifiltered ring for $p = 3, 5, 7$, and in all cases we have found consistency with the Macdonald index for all the coefficients we have been able to verify in a reasonable time. To illustrate this, we write here the Hilbert series for the case of $p = 5$:

\vspace{-12pt}
\scriptsize
\begin{equation*}
\begin{aligned}
I_M{}&{}^{\mathcal{D}_5(SU(4))}=1+q(15T) +q^2(16T+119T^2) +q^3(16T+240T^2+664T^3) +q^4(16T+376T^2+1904T^3+2924T^4)\\
&+q^5(16T+496T^2+3944T^3+10608T^4+10814T^5) +q^6(16T+632T^2+6560T^3+26791T^4+46560T^5+34916T^6)\\
&+q^7(16T+752T^2+9960T^3+53311T^4+136593T^5+171360T^6+101048T^7)\\
&+q^8(16T+888T^2+13936T^3+93771T^4+313105T^5+565321T^6+549936T^7+267137T^8)\\
&+q^9(16T+1008T^2+18712T^3+149871T^4+623117T^5+1451990T^6+1993403T^7+1580592T^8+654381T^9)\\
&+q^{10}\bigl(16T+1144T^2+24048T^3+225591T^4+1114673T^5+3205866T^6+5617722T^7+6187422T^8+4147728T^9\\
&\qquad\quad{}+1501799T^{10}\bigr)+O(q^{11}) \,.
\end{aligned}
\end{equation*}
\normalsize

\paragraph{\boldmath{$\mathcal{D}_p(SU(5))$}:}

Finally, we turn to the $\mathcal{D}_p(SU(5))$ theories. We find the potentials are as given in Table \ref{tab:DpSUN}. These are written in terms of the Casimir invariants of $\mathfrak{su}(5)$, which are polynomials in the twenty-four generators of the adjoint representation. As before, we introduce additional generators with non-homogeneous linear relations to capture the $T$-grading necessary for the bifiltration structure.
We have
\begin{align}
    t = C_2(\vec{a}) \ , \quad  w_3 = C_3(\vec{a}) \ , \quad w_4 = C_4(\vec{a}) \ , \quad w_5 = C_5 (\vec{a}) \,,
\end{align} 
with $T$-grading $T(t)=1, T(w_3)=2, T(w_4)=3, T(w_5) = 4$. The proposed bifiltered ring is then clear. For $p = 6$, the Hilbert series of the arc space is:\footnote{We have also computed this quantity for $p = 2, 3, 4$, and found agreement with the Macdonald index.}

\vspace{-12pt}
\scriptsize
\begin{equation*}
\begin{aligned}
I_M{}&{}^{\mathcal{D}_6(SU(5))}=1+q(24T) +q^2(25T+299T^2) +q^3(25T+600T^2+2575T^3) +q^4(25T+925T^2+7475T^3+17225T^4)\\
&+q^5(25T+1225T^2+15275T^3+64375T^4+95356T^5) +q^6(25T+1550T^2+25400T^3+161550T^4+430600T^5+454571T^6)\\
&+q^7(25T+1850T^2+38400T^3+321450T^4+1267474T^5+2383325T^6+1917396T^7)\\
&+q^8\bigl(25T+2175T^2+53725T^3+563900T^4+2914549T^5+7980801T^6+11357400T^7+7298224T^8\bigr)\\
&+q^9\bigl(25T+2475T^2+71950T^3+901250T^4+5797624T^5+20679526T^6+42333426T^7+47878000T^8+25442924T^9\bigr)\\
&+q^{10}\bigl(25T+2800T^2+92475T^3+1354250T^4+10381854T^5+45789026T^6+121300251T^7+195439125T^8\\
&\qquad\quad{}+182089350T^9+82184544T^{10}\bigr)+O(q^{11}) \,,
\end{aligned}
\end{equation*}
\normalsize
which is consistent with both the Schur and Hall--Littlewood indices of the $\mathcal{D}_6(SU(5))$ Argyres--Douglas SCFT. The Macdonald index, which was heretofore unknown, is therefore predicted to begin with this expansion.

\section{Geometrizing renormalization group flows}\label{sec:FLOWS}

In Section \ref{sec:RINGS}, we determined the bifiltered rings associated to a collection of Argyres--Douglas SCFTs using syzygy maximization, and verified that these rings reproduce the Macdonald index (or the known limits) according to equation \eqref{eqn:bigMAC}. Many of the Argyres--Douglas theories we studied are related by Higgs branch renormalization group flows. In Section \ref{sec:conj}, we made a conjecture for how the bifiltered ring is modified under Higgs branch RG flow, in particular for RG flow via nilpotent Higgsing of the moment map of a simple non-Abelian global symmetry, where the only modes to decouple along the flow are Nambu--Goldstone modes from the moment map. In this section, we show that the conjecture of Section \ref{sec:conj} holds for all such RG flows between the SCFTs studied in Section \ref{sec:RINGS}.

\subsection{Higgsing \texorpdfstring{$\mathcal{D}_p(SU(2))$}{Dp(SU2)} to \texorpdfstring{$\mathcal{D}_p(SU(2),[2]) = (A_1, A_{p - 3})$}{Dp(SU2,[2]) = (A1, Ap-3)}}

Let us start by considering a particularly simple family of examples: the Argyres--Douglas theories $\mathcal{D}_{2n + 1}(SU(2))$. The universal polynomial, which is of degree $p + 1 = 2n + 2$, can only consist of a single term:
\begin{equation}
    P_u(u_2) = u_2^{n+1} \,,
\end{equation}
and there are no coefficients to be fixed by imposing syzygy maximization. Therefore, our bifiltered ring is derived from the Jacobian ideal of the polynomial
\begin{equation}
    P_{[1^2]}(x,y,z) = P_u(C_2(x,y,z)) = (xy + z^2)^{n+1} \,.
\end{equation}
The ideal that we would determine from equation \eqref{eqn:ringgen} is
\begin{equation}
    \left(x(xy + z^2)^n, y(xy + z^2)^n, z(xy + z^2)^n, w - (xy + z^2) \right) \,.
\end{equation}
Since $w$ has a lower $T$-grading than $(xy + z^2)$, we can write an equivalent ideal for the bifiltered ring as
\begin{equation}
    \left(xw^n, yw^n, zw^n, w - (xy + z^2) \right) \,.
\end{equation}
This is precisely the ideal that was conjectured in \cite{Kang:2025zub,Andrews:2025krn}.

Now, we can consider the Higgsing of the $\mathfrak{su}(2)$ global symmetry by giving a VEV valued in the nilpotent orbit associated to the partition $[2]$. The Slodowy slice, $\mathcal{S}_{[2]}$, has only a single affine coordinate, $t$, such that
\begin{equation}
    C_2|_{\mathcal{S}_{[2]}}(t) = t \,.
\end{equation}
Thus, we obtain the bifiltered ring
\begin{equation}
  \begin{gathered}
    \mathbb{C}[t,w] \, / \, \left( t^n, w - t \right) \,, \qquad
    \operatorname{deg}(t) = (2,2) \,, \quad \operatorname{deg}(w) = (2,1) \,.
  \end{gathered}
\end{equation}
Since $w$ has a lower $T$-degree than $t$, the above is equivalent as a bifiltered quotient ring to
\begin{equation}\label{eqn:pooh}
    \mathbb{C}[w] \, / \, \left( w^n \right) \,, \qquad \operatorname{deg}(w) = (2,1) \,.
\end{equation}
In this case, the bifiltration is actually a bigrading as the ideal is generated by a polynomial which is homogeneous in both the $q$-grading and the $T$-grading. The bigraded ring in equation \eqref{eqn:pooh} is precisely the ring associated to the Macdonald index and Higgs branch of the $(A_1, A_{2n-2})$ Argyres--Douglas theory in \cite{Bhargava:2023hsc}.

Thus, we have shown that the general procedure for tracking the bifiltered rings under Higgs branch RG flows, conjectured in Section \ref{sec:conj}, holds for the infinite family of flows:
\begin{equation}
    \mathcal{D}_{2n+1}(SU(2)) = (A_1, D_{2n+1}) \quad \xrightarrow{\,\, a_1 \,\,} \quad  \mathcal{D}_{2n+1}(SU(2), [2]) = (A_1, A_{2n-2}) \,. 
\end{equation}

\subsection{Higgsing \texorpdfstring{$\mathcal{D}_{11}(SU(3))$}{D11(SU3)} to \texorpdfstring{$\mathcal{D}_{11}(SU(3), [2,1])$}{D11(SU3, [2,1])} to \texorpdfstring{$\mathcal{D}_{11}(SU(3), [3])$}{D11(SU3, [3])}}\label{sec:D11eg}

We now turn to a simple, explicit example where we can see the syzygy maximization principle in action, to fix the coefficients in the ideals defining the bifiltered rings, and how the property of syzygy maximization is preserved under Higgs branch RG flow.

We
consider the Argyres--Douglas theory $\mathcal{D}_{11}(SU(3))$, and the other
two interacting SCFTs that exist on subloci of its Higgs branch.  The three
SCFTs in question are related via the following sequence of Higgs branch RG
flows: 
\begin{equation}\label{eqn:RGexample}
    \mathcal{D}_{11}(SU(3)) \,\,\xrightarrow{\,\,a_2\,\,} \,\,\mathcal{D}_{11}(SU(3), [2,1]) \,\,\xrightarrow{\,\,A_2\,\,} \,\, \mathcal{D}_{11}(SU(3), [3]) = (A_2, A_7) \,.
\end{equation}
The labels on the arrows indicate the elementary slices in the Higgs branch, as
a symplectic singularity; $a_2$ is the closure of the minimal nilpotent orbit
of $\mathfrak{su}(3)$ and $A_2$ is the orbifold singularity
$\mathbb{C}^2/\mathbb{Z}_3$. The Higgs branch is the closure of the principal
nilpotent orbit of $\mathfrak{su}(3)$.

The bifiltered ring $R$ underlying the Macdonald index for
$\mathcal{D}_{11}(SU(3))$ was derived in Section \ref{sec:DpSUN} using the principle of syzygy maximization, and is as
follows. Consider the bigraded polynomial ring
\begin{equation}
  \begin{gathered}
    S_{[1^3]} = \mathbb{C}[a_1, \cdots, a_8, t, w] \,, \\ \operatorname{deg}(a_i) = (1, 1) \,,\,\,\, \operatorname{deg}(t) = (2, 1) \,, \,\,\, \operatorname{deg}(w) = (3, 2)  \,.
  \end{gathered}
\end{equation}
The $a_i$ parametrize an element of the adjoint representation of $\mathfrak{su}(3)$, that is
\begin{equation}\label{eqn:su3el}
    X = \sum_{i=1}^8 a_i T_i \,,
\end{equation}
where $T_i$ are the $\mathfrak{su}(3)$ generators. Since $\mathfrak{su}(3)$ is
a global symmetry of the theory, we take the potential to be $\mathfrak{su}(3)$ invariant. Therefore, it is a polynomial in the
invariant polynomials of $\mathfrak{su}(3)$.\footnote{For ease of notation, we
do not distinguish between $\mathfrak{su}(3)$ and $\mathfrak{sl}(3)$ explicitly in this
paper, but we assume the reader can make the appropriate reading in each case.} The $\mathfrak{su}(3)$-invariant polynomial ring is freely-generated by
the degree two and three Casimir invariants of $\mathfrak{su}(3)$, that is
$\mathbb{C}[C_2(\bm{a}), C_3(\bm{a})]$, where $C_k(\bm{a})$ denotes the $k$th
Casimir invariant of $\mathfrak{su}(3)$. Specifically,
\begin{equation}
  \begin{aligned}
    C_2(\bm{a}) = \frac{1}{2}\operatorname{tr}X^2 \,, \qquad
    C_3(\bm{a}) = \frac{1}{3}\operatorname{tr} X^3 \,, 
  \end{aligned}
\end{equation}
where $X$ is the $\mathfrak{su}(3)$ element as in equation \eqref{eqn:su3el}.
Then, we can define the following degree twelve $\mathfrak{su}(3)$-invariant polynomial:
\begin{equation}\label{eqn:P13}
    P_{[1^3]}(C_2, C_3) = C_3^4 + \frac{2}{3} C_3^2 C_2^3 + \frac{2}{135} C_2^6   \,.
\end{equation}
We consider a Jacobian ideal associated to $P_{[1^3]}$, augmented by linear
relations for the coordinates $t$ and $w$:
\begin{equation}\label{eqn:I13}
    I_{[1^3]} = \left( \partial_{a_1} P_{[1^3]} ,\, \cdots ,\, \partial_{a_8} P_{[1^3]}, 
    t - C_2(\bm{a}), w - C_3(\bm{a}) \right) \,.
\end{equation}
The bifiltered ring associated to $\mathcal{D}_{11}(SU(3))$ is then
\begin{equation}\label{eqn:13ring}
    R_{[1^3]} = S_{[1^3]} \,/\, I_{[1^3]} \,,
\end{equation}
where the bifiltration is that induced from the bigrading on $S_{[1^3]}$. The
coefficients appearing in equation \eqref{eqn:P13} are determined by imposing a
syzygy maximization condition of the form written in equation
\eqref{eqn:smaxgen} to the Jacobian ideal in equation
\eqref{eqn:I13}.\footnote{We can solve for the linear relations for $t$ and $w$
to obtain a ring $R$ which is obtained by quotienting by an honest Jacobian
ideal, however, this obscures the structure of the bifiltration, which cannot
then be inherited from a bigrading on the polynomial ring, and thus we choose
to keep the additional coordinates and linear relations. We want to emphasize
that this is a point of presentation only: the ring after solving the linear
relations is isomorphic to the ring defined here, and the bifiltration can be
pushed through the isomorphism.}

Similarly, for $\mathcal{D}_{11}(SU(3), [2,1])$, the bifiltered ring was
determined in Section \ref{sec:DpSUnRG}. Again, we start with a bigraded
polynomial ring 
\begin{equation}
  \begin{gathered}
    S_{[2,1]} = \mathbb{C}[x,y,z,t,w] \,, \\[0.1em] \operatorname{deg}(x) = \operatorname{deg}(y) = \left(\frac{3}{2}, \frac{3}{2}\right) \,,\,\,\, \operatorname{deg}(z) = (1,1)\,,\,\,\, \operatorname{deg}(t) = (2,1)\,,\,\,\, \operatorname{deg}(w) = (3,2) \,.
  \end{gathered}
\end{equation}
Analogously to the introduction of the Casimirs for $\mathcal{D}_{11}(SU(3))$,
we start by defining an intermediate polynomial, which is
\begin{equation}
    \widetilde{w}(x,y,z,t) = xy + z^3 + zt \,,
\end{equation}
and a degree twelve polynomial
\begin{equation}
    P_{[2,1]}(\widetilde{w}, t) = \widetilde{w}^4 - \frac{2}{3}\widetilde{w}^2 t^3 + \frac{2}{135} t^6 \,.
\end{equation}
We then consider the associated Jacobian ideal, augmented by adding a linear relation for $w$ which fixes the bifiltered structure:
\begin{equation}
    I_{[2,1]} = \left( \partial_x P_{[2,1]} ,\,  \partial_y P_{[2,1]} ,\,  \partial_z P_{[2,1]} ,\, \partial_t P_{[2,1]} ,\, w - \widetilde{w}\right) \,.
\end{equation}
Therefore, the bifiltered ring underlying both the Higgs branch and the Macdonald index for the $\mathcal{D}_{11}(SU(3), [2,1])$ SCFT is
\begin{equation}\label{eqn:21ring}
    R_{[2,1]} = S_{[2,1]} \,/\, I_{[2,1]} \,.
\end{equation}

Finally, for $(A_2, A_7)$, the bifiltered ring was determined previously in \cite{Kang:2026nge}. We start with the bigraded polynomial ring
\begin{equation}
    S_{[3]} = \mathbb{C}[x, y] \,, \qquad \operatorname{deg}(x) = (3,2) \,, \quad \operatorname{deg}(y) = (2,1) \,,
\end{equation}
and the ideal is a Jacobian ideal, which we refer to as $I_{[3]}$, derived from the potential
\begin{equation}\label{eqn:3pot}
    P_{[3]}(x,y) = x^4 + x^2 y^3 + \frac{1}{30} y^6 \,.
\end{equation}

Now that we have recapped the bifiltered rings for each of the SCFTs
appearing in the Higgs branch RG flows in equation \eqref{eqn:RGexample}, we
are ready to discuss the geometric realization of the RG flows in terms of the
bifiltered rings. First, we observe that we can define a universal potential
\begin{equation}\label{eqn:universalpoly}
    P(u, v) = v^4 + \frac{2}{3} u^3 v^2 + \frac{2}{135} u^6 \,,
\end{equation}
such that 
\begin{equation}
    P_{[1^3]} = P(C_2(a), C_3(a)) \,, \quad P_{[2,1]} = P(-t, xy + z^3 + zt) \,, \quad P_{[3]} = P(y, x) \,,
\end{equation}
up to an overall rescaling of the coordinates (which leaves the Jacobian ideal
invariant). A priori, this is surprising: the coefficients in $P_O$ were fixed
by studying the algebraic Betti numbers of the Jacobian ideal obtained by
taking (non-trivial) derivatives with respect to, respectively, eight, four,
and two coordinates. Regardless of this difference, we find that the
syzygy maximization condition in equation \eqref{eqn:smaxgen} leads to a
unique, universal set of coefficients in the potentials.

The fact that these three bifiltered rings can all be derived from the same
universal potential should come as no surprise, after the reader has become
familiar with the conjecture in Section \ref{sec:conj}. The three polynomials,
$P_O$, given above were determined via algebro-geometric
bootstrapping/syzygy maximization, a determination which is independent of the
structure of the Higgs branch as a symplectic singularity. We now explain how
the three rings are consistent with the Higgs branch structure in equation
\eqref{eqn:RGexample}.

The coordinates $a_i$ in the polynomial ring $S_{[1^3]}$ correspond to the
$\widehat{B}_1$ moment-map operators of the 4d $\mathcal{D}_{11}(SU(3))$
theory.\footnote{For the representation theory of the 4d superconformal
algebras, we follow the notation of \cite{Dolan:2002zh}.} This four-dimensional
information fixes the $(q,T)$-bigrading of the $a_i$. In fact, these $a_i$
transform in the adjoint representation of the $\mathfrak{su}(3)$ global
symmetry. Under nilpotent Higgsing of the $\mathfrak{su}(3)$ symmetry, such as
the Higgsing appearing in equation \eqref{eqn:RGexample}, it is well-known
(see, for example,
\cite{Tachikawa:2015bga,Distler:2022nsn,Beem:2023ofp,Couzens:2023kyf,Baume:2023onr})
what happens to the moment-map operators; in particular, which of them decouple
and become free along the RG flow, and how they are charged under the infrared
R-symmetry.

Let $O$ be a nilpotent orbit of $\mathfrak{su}(3)$ and choose an
associated $\mathfrak{sl}(2)$-triple, $(e, h, f)$, via the Jacobson--Morozov theorem.\footnote{For the standard conventions for nilpotent orbits, we
refer to the canonical work \cite{MR1251060}.} The Slodowy slice associated to
the nilpotent orbit is formally:
\begin{equation}
    \mathcal{S}_O = e + \operatorname{ker}(\operatorname{ad} f) \,.
\end{equation}
The slice $\mathcal{S}_O$ is parametrized by a collection of affine coordinates, $n_1, \cdots,
n_{d}$, which we refer to collectively as $\bm{n}$. There is also a natural
grading, called the (half-)Kazhdan grading on the Slodowy slice, and we write the
charge of the $n_i$ under this grading as $q_i$. Then, we consider the
following bigraded polynomial ring
\begin{equation}
    S_O = \mathbb{C}[\bm{n}] \otimes_{\mathbb{C}} \mathbb{C}[t, w] \,, \qquad \operatorname{deg}(n_i) = (q_i, q_i) \,, \quad \operatorname{deg}(t) = (2,1) \,, \quad \operatorname{deg}(w) = (3, 2) \,.
\end{equation}
We now define a polynomial belonging to $S_O$ as follows:
\begin{equation}
    P_{O}(\bm{n}) = P(C_2|_{\mathcal{S}_O}(\bm{n}), C_3|_{\mathcal{S}_O}(\bm{n})) \,,
\end{equation}
where $P$ is the universal polynomial we have defined in equation
\eqref{eqn:universalpoly}, and the Casimir invariants restricted to the Slodowy
slice associated to the nilpotent orbit are 
\begin{equation}
    C_2|_{\mathcal{S}_O}(\bm{n}) = \frac{1}{2} \operatorname{tr} X^2 \,, \quad C_3|_{\mathcal{S}_O}(\bm{n}) = \frac{1}{3} \operatorname{tr} X^3 \,,
\end{equation}
where $X$ is the generic element of the Slodowy slice, which is parametrized by
the affine coordinates on the slice, $\bm{n}$. We then propose, as stated generically in Section \ref{sec:conj}, that the
bifiltered ring associated to the $\mathcal{D}_{11}(SU(3), O)$ is
\begin{equation}\label{eqn:THESU3RING}
    S_O \, / \, \left( \partial_{n_1} P_O(\bm{n}), \cdots, \partial_{n_d} P_O(\bm{n}), t - C_2|_{\mathcal{S}_O}(\bm{n}), w - C_3|_{\mathcal{S}_O}(\bm{n}) \right) \,.
\end{equation}
In this way, we see clearly how the bifiltered ring transforms under the Higgs
branch RG flows generated by giving a nilpotent vacuum expectation value to the
moment map of the $\mathfrak{su}(3)$ global symmetry of
$\mathcal{D}_{11}(SU(3))$.

We now look at this explicitly for the theories in equation
\eqref{eqn:RGexample}. For $\mathcal{D}_{11}(SU(3))$, it is trivial to see that
the ring in equation \eqref{eqn:THESU3RING} is the same as that in equation
\eqref{eqn:13ring}, once we note that the $n_i$ are simply the $a_i$ appearing
in equation \eqref{eqn:su3el}. For the nilpotent orbit associated to the
$[2,1]$ partition, the Slodowy slice can be parametrized by matrices of the
form
\begin{equation}
    X = \begin{pmatrix}
      -\frac{n_4}{2} & 1 & 0 \\
      -n_1 - \frac{3}{4}n_4^2 & -\frac{n_4}{2} & n_2 \\
      n_3 & 0 & n_4 
    \end{pmatrix} \,,
\end{equation}
in terms of the four affine coordinates $\bm{n}$. Then, we can see that the Casimirs restricted to the slice become
\begin{equation}\label{eqn:thomas}
    C_2|_{\mathcal{S}_{[2,1]}}(\bm{n}) = -n_1 \,, \qquad C_3|_{\mathcal{S}_{[2,1]}}(\bm{n}) = -n_1 n_4 - n_2 n_3 - n_4^3 \,.
\end{equation}
It is then straightforward to see that the ring associated to this nilpotent
orbit as in equation \eqref{eqn:THESU3RING} is the same as that we derived
explicitly in equation \eqref{eqn:21ring}. For the $[3]$ partition, the Slodowy
slice has two coordinates, $n_1$ and $n_2$, with half-Kazhdan grading
\begin{equation}
    q_1 = 2 \,, \quad q_2 = 3 \,,
\end{equation}
and
\begin{equation}
    C_2|_{\mathcal{S}_{[3]}}(\bm{n}) = n_1 \,, \qquad C_3|_{\mathcal{S}_{[3]}}(\bm{n}) = n_2 \,.
\end{equation}
Explicitly, the ring that we would obtain following from equation \eqref{eqn:THESU3RING} is:
\begin{equation}
  \begin{gathered}
    \mathbb{C}[n_1, n_2, t, w] \, / \, \left( \partial_{n_1} P(n_1, n_2) \,, \partial_{n_2} P(n_1, n_2) \,, t - n_1 \,, w - n_2  \right) \\
    \operatorname{deg}(n_1) = (2, 2) \,, \,\,\operatorname{deg}(n_2) = (3, 3) \,, \,\,\operatorname{deg}(t) = (2, 1) \,, \,\,\operatorname{deg}(w) = (3, 2) \,.
  \end{gathered}
\end{equation}
Solving for the linear relations leads to a bifiltered ring which is isomorphic
to that we wrote for the $(A_2, A_7)$ theory derived from the potential in
equation \eqref{eqn:3pot}.

\subsection{Higgsing \texorpdfstring{$\mathcal{D}_p(SU(3))$}{Dp(SU(3))} to \texorpdfstring{$\mathcal{D}_p(SU(3), O)$}{Dp(SU(3), O)}}

In the previous subsection, we studied the Higgs branch RG flows from the $\mathcal{D}_{11}(SU(3))$ SCFT. However, in Section \ref{sec:RINGS}, we determined the bifiltered rings not just for $p = 11$. We found the rings for the theories
\begin{equation}
  \begin{aligned}
    &\mathcal{D}_p(SU(3)) & &\qquad \text{ for } p = 2, 4, 5, 7, 8, 10, 11 \,, \\
    &\mathcal{D}_p(SU(3), [2,1]) & &\qquad \text{ for } p = 4, 5, 7, 8, 10, 11 \,.
  \end{aligned}
\end{equation}
Furthermore, in \cite{Kang:2026nge}, the bifiltered rings were determined for
\begin{equation}\label{eqn:cashews}
    \mathcal{D}_p(SU(3), [3]) \qquad \text{ for } p = 5, 7, 8, 10, 11, 13, 14, 16, 17, 19, 20, 22, 23, 25, 26 \,,
\end{equation}
in their guise as the $(A_2, A_{p-4})$ theories. 

For any $p > 2$, such that $\gcd(p, 3) = 1$, there exists the natural generalization of the Higgs branch RG flow in equation \eqref{eqn:RGexample}, to wit:
\begin{equation}\label{eqn:salt}
    \mathcal{D}_{p}(SU(3)) \,\,\xrightarrow{\,\,a_2\,\,} \,\,\mathcal{D}_{p}(SU(3), [2,1]) \,\,\xrightarrow{\,\,A_2\,\,} \,\, \mathcal{D}_{p}(SU(3), [3]) \,.
\end{equation}
For $p = 4$, the final theory in this Higgs branch RG flow is the trivial theory, which is why we have not counted the bifiltered ring explicitly in equation \eqref{eqn:cashews}. For $p = 2$, the flow truncates early:
\begin{equation}\label{eqn:pepper}
    \mathcal{D}_{2}(SU(3)) \,\,\xrightarrow{\,\,a_2\,\,} \,\,\mathcal{D}_{2}(SU(3), [2,1]) \,,
\end{equation}
where the final theory is again the trivial theory. 

Since we have determined the bifiltered rings independently of the structure of the Higgs branch, using the principle of syzygy maximization, we can then ask whether the rings are consistent with the Higgsing conjecture in Section \ref{sec:conj} along the flows in equations \eqref{eqn:salt} or \eqref{eqn:pepper}. For each $2 \leq p < 11$ such that $\gcd(p, 3) = 1$, a straightforward modification of the analysis of Section \ref{sec:D11eg} reveals that the rings derived from syzygy maximization are equivalent to those from the conjecture in Section \ref{sec:conj}.

\subsection{Higgsing \texorpdfstring{$\mathcal{D}_p(SU(4))$}{Dp(SU4)} to \texorpdfstring{$\mathcal{D}_p(SU(4), O)$}{Dp(SU4,O)}}

In Section \ref{sec:RINGS}, we have determined the bifiltered rings that underlie the Higgs branch and Macdonald index for the theories:
\begin{equation}
    \begin{aligned}
        &\mathcal{D}_p(SU(4)) & &\qquad \text{ for } p = 3, 5, 7 \,, \\
        &\mathcal{D}_7(SU(4), [3,1]) \,.
    \end{aligned}
\end{equation}
Furthermore, in \cite{Kang:2026nge}, we determined the bifiltered rings for the theories
\begin{equation}
    \mathcal{D}_p(SU(4), [4]) = (A_{p-5}, A_3) \qquad \text{ for } p = 7, 9, 11, 13, 15 \,.
\end{equation}
Therefore, in this section, we study the reflection of the Higgs branch RG flow:
\begin{equation}\label{eqn:frenchtoast}
    \mathcal{D}_7(SU(4)) \,\,\longrightarrow\,\, \mathcal{D}_7(SU(4), [3,1]) 
    \,,
\end{equation}
in the bifiltered rings that we have determined are associated to each of the two theories. We study some examples of Higgsings from $\mathcal{D}_p(SU(4))$ where $p = 5$ in Section \ref{sec:genericpNp1} and $p = 3$ in Section \ref{sec:isos}. We note that the arrow in equation \eqref{eqn:frenchtoast} is not a minimal Higgsing, in fact it represents the following sequence of Higgsings:
\begin{equation}
    \mathcal{D}_7(SU(4)) \xrightarrow{\,\,a_3\,\,} \mathcal{D}_7(SU(4), [2, 1^2]) \xrightarrow{\,\,a_1\,\,} \mathcal{D}_7(SU(4), [2^2]) \xrightarrow{\,\,a_1\,\,} \mathcal{D}_7(SU(4), [3,1])  \,.
\end{equation}
Since we understand the bifiltered ring for $\mathcal{D}_7(SU(4))$, we can use the conjecture in Section \ref{sec:conj} to determine the bifiltered rings for each of the theories in this sequence.

We consider the nilpotent Higgsing associated to the $[3,1]$ partition. After the Higgsing, the Higgs branch is given by the intersection of the nilpotent cone $\mathcal{N}_{\mathfrak{sl}(4)}$ with the corresponding Slodowy slice $\mathcal{S}_{[3, 1]}$ of $\mathfrak{sl}(4)$. The Slodowy slice has 5 coordinates: we label these coordinates as $n_1, \cdots, n_5$, and the generic $\mathfrak{su}(4)$ matrix belonging to the Slodowy slice, after picking a basis, can be written as
\begin{align}
 M = \left( \begin{array}{cccc}
    -n_1 & 1 & 0 & 0 \\
    n_2 & -n_1 & 1 & 0 \\
    n_5 & n_2 & -n_1 & n_3 \\
    n_4 & 0 & 0 & 3n_1 
 \end{array} \right) \,. 
\end{align}
It is then straightforward to determine the Casimir invariants restricted to the Slodowy slice using the trace formulae:
\begin{equation}\label{eqn:su431Cas}
  \begin{aligned}
    C_2|_{\mathcal{S}_{[3,1]}}(\bm{n}) &= 2(3n_1^2 + n_2) \,,\\
    C_3|_{\mathcal{S}_{[3,1]}}(\bm{n}) &= 8n_1^3 - 4n_1n_2 + n_5 \,,\\
    C_4|_{\mathcal{S}_{[3,1]}}(\bm{n}) &= 21n_1^4 + 6n_1^2n_2 - 3n_1n_5 + 2n_2^2 + n_3n_4 \,.
  \end{aligned}
\end{equation}
We can reparametrize the affine coordinates on the Slodowy slice as follows:
\begin{equation}
    t = -2(3n_1^2 + n_2) \,, \quad w = 8n_1^3 - 4n_1n_2 + n_5 \,, \quad z = -3n_1 \,, \quad x = n_3 \,, \quad y = n_4 \,.
\end{equation}
then we have
\begin{equation}
  \begin{aligned}
    C_2|_{\mathcal{S}_{[3,1]}}(x,y,z,t,w) &= -t \,, \\
    C_3|_{\mathcal{S}_{[3,1]}}(x,y,z,t,w) &= w \,, \\
    C_4|_{\mathcal{S}_{[3,1]}}(x,y,z,t,w) &= xy + z^4 + tz^2 + wz + \frac{t^2}{2} \,, \\
  \end{aligned}
\end{equation}

The intersection with the nilpotent cone $\mathcal{N}_{\mathfrak{sl}(4)} \cap \mathcal{S}_{[3, 1]}$ is given by $C_2=C_3=C_4=0$, which gives the Kleinian singularity $\mathbb{C}^2/\mathbb{Z}_4 = \{xy+z^4=0 \}$. 
Plugging these restricted Casimirs into equation for $p=7$ and $N=4$ in Table \ref{tab:DpSUN} indeed reproduces equation \eqref{eq:D7su4Higgs}, where these polynomials are obtained by independent syzygy maximization. Therefore, we see again that the RG flow can be implemented at the level of the geometric potential.

\subsection{Higgsing \texorpdfstring{$\mathcal{D}_6(SU(5))$}{D6(SU(5))} to \texorpdfstring{$\mathcal{D}_6(SU(5), [4,1])$}{D6(SU(5), [4,1])}}

We now consider the $\mathcal{D}_6(SU(5))$ SCFT and some of its Higgs branch RG flows. The generic element of $\mathfrak{su}(5)$ is a linear combination of $24$ generators, and thus the underlying polynomial ring contains $24$ variables. In Section \ref{sec:RINGS}, we were able to determine the bifiltered ring for $p = 6$, by performing syzygy maximization across the $48$ generators of the polynomial ring underlying the first jet scheme. In principle, we could perform the same process for any $\mathcal{D}_6(SU(5), O)$, that is, for any nilpotent orbit belonging to $\mathfrak{su}(5)$. There are seven such nilpotent orbits, and under the standard dominance ordering, they can be written as:
\begin{equation}
[1^5] \,\,\rightarrow\,\, [2, 1^3] \,\,\rightarrow\,\, [2^2, 1] \,\,\rightarrow\,\, [3, 1^2] \,\,\rightarrow\,\, [3,2] \,\,\rightarrow\,\, [4,1] \,\,\rightarrow\,\, [5] \,.
\end{equation}
For $p = 6$, using the principle of syzygy maximization, we have determined the bifiltered rings for only two theories in the RG flow:
\begin{equation}
    \mathcal{D}_6(SU(5)) \qquad \text{ and } \qquad \mathcal{D}_6(SU(5), [4,1]) = (A_1, A_9) \,.
\end{equation}
Since we have the ring for the former, we have the universal potential which is a necessary input to the conjecture in Section \ref{sec:conj}, and thus we can conjecture the bifiltered rings for any $\mathcal{D}_6(SU(5), O)$. We can provide a further verification of the conjecture by checking that the conjecture for the $[4,1]$ Higgsing matches with the derivation for the $(A_1, A_9)$ theory. 

We recall the universal potential for the $\mathcal{D}_6(SU(5), O)$ family of theories as derived from the principle of syzygy maximization in Section \ref{sec:RINGS}:
\begin{equation}
    P(u_2, \cdots, u_5) = u_5 u_2 + u_4 u_3 - \frac{11}{10}u_3u_2^2 \,.
\end{equation}
For $O = [4,1]$, the Slodowy slice is six-dimensional, and the half-Kazhdan degrees of the coordinates are
\begin{equation}\label{eqn:gouda}
    (\operatorname{deg}(n_1), \cdots, \operatorname{deg}(n_{6})) = \left( 1, 2, \frac{5}{2}, \frac{5}{2}, 3, 4 \right) \,.
\end{equation}
The restrictions of the Casimirs to this Slodowy slice are
\begin{equation}
    \begin{aligned}
        C_2|_{\mathcal{S}_{[4,1]}} &= 10n_1^2+10n_2  \,,\\
        C_3|_{\mathcal{S}_{[4,1]}} &= 20n_1^3-20n_1n_2+2n_5  \,,\\
        C_4|_{\mathcal{S}_{[4,1]}} &= 65n_1^4+30n_1^2n_2-6n_1n_5+41n_2^2+n_6 \,,\\
        C_5|_{\mathcal{S}_{[4,1]}} &= 204n_1^5-40n_1^3n_2+12n_1^2n_5-164n_1n_2^2 -4n_1n_6+20n_2n_5+n_3n_4 \,.
    \end{aligned}
\end{equation}
Now, we can evaluate the universal polynomial pulled back to the polynomial  ring of the Slodowy slice:
\begin{equation}
  \begin{aligned}
    P(\bm{n}) &= P_u(C_2|_{\mathcal{S}_{[4,1]}}, C_3|_{\mathcal{S}_{[4,1]}}, C_4|_{\mathcal{S}_{[4,1]}}, C_5|_{\mathcal{S}_{[4,1]}}) \,.
  \end{aligned}
\end{equation}
We find that the Jacobian ideal contains homogeneous linear relations, and thus it is equivalent, after eliminating $n_5,n_6$ and making homogeneous changes of the remaining coordinates, whose relabeling we suppress, to the following:
\begin{equation}
    \left( n_2 n_3, n_2 n_4, n_1^5 -20n_1^3n_2 + 30 n_1 n_2^2 + n_3 n_4, n_2(n_1^4 - 6n_1^2n_2 + 2n_2^2) \right)\,,
\end{equation}
where this is an ideal belonging to the polynomial ring $\mathbb{C}[n_1, \cdots, n_4]$ where the bidegrees are given via the half-Kazhdan grading as in equation \eqref{eqn:gouda}. When quotienting by this ideal, the restricted Casimirs can be written as
\begin{equation}
    \begin{aligned}
        C_2|_{\mathcal{S}_{[4,1]}} &= n_2  \,,\\
        C_3|_{\mathcal{S}_{[4,1]}} &= n_1 n_2 \,,\\
        C_4|_{\mathcal{S}_{[4,1]}} &= n_2(2 n_1^2 + n_2) \,,\\
        C_5|_{\mathcal{S}_{[4,1]}} &= n_1n_2 (n_1^2 + n_2)\,.
    \end{aligned}
\end{equation}
Therefore, the ideal of interest is
\begin{equation}
  \begin{aligned}
     \big( &n_2 n_3, n_2 n_4, n_1^5 -20n_1^3n_2 + 30 n_1 n_2^2 + n_3 n_4, n_2(n_1^4 - 6n_1^2n_2 + 2n_2^2), \\&\qquad\qquad w_2 - n_2, w_3 - n_1 n_2, w_4 - n_2(2n_1^2 + n_2), w_5 - n_1 n_2(n_1^2 + n_2) \big)\,.
  \end{aligned}
\end{equation}
We can solve for the linear relations for $n_2$, $w_3$, $w_4$, and $w_5$, since they have the largest $T$-grading in their respective relations. We relabel as 
\begin{equation}
    n_1 \rightarrow z \,, \quad w_2 \rightarrow t \,, \quad n_3 \rightarrow x \,, \quad n_4 \rightarrow y \,,
\end{equation}
and thus we end up with the ideal:
\begin{equation}\label{eqn:sujebi}
     \big( t x, t y, z^5 -20z^3t + 30 z t^2 + x y, t(z^4 -6z^2t + 2t^2) \big) \,,
\end{equation}
belonging to the ring 
\begin{equation}\label{eqn:garlic}
    \mathbb{C}[x,y,z,t] \,, \qquad \operatorname{deg}(x) = \operatorname{deg}(y) = \left(\frac{5}{2}, \frac{5}{2}\right) \,, \,\,\operatorname{deg}(z) = (1,1) \,, \,\,\operatorname{deg}(t) = (2,1) \,.
\end{equation}

The bifiltered ring associated to the $(A_1, A_9)$ SCFT was determined in Section \ref{sec:RINGS} via syzygy maximization. The potential belonged to the same polynomial ring as written in equation \eqref{eqn:garlic} and was given as
\begin{equation}
    P(x,y,z,t) = 
    (xy + z^5)t + z^3t^2 + \frac{1}{10} zt^3 \,.
\end{equation}
Computing the Jacobian ideal, we find
\begin{equation}
    \left( xt, yt, (xy + z^5) + 2 z^3 t + \frac{3}{10}zt^2, t(5z^4 + 3 z^2 t + \frac{1}{10}t^2) \right) \,.
\end{equation}
Under the rescaling $t \rightarrow -10t$, this is exactly identical to the ideal that we derived from the $[4,1]$ partition in equation \eqref{eqn:sujebi}. Therefore, we have shown that the bifiltered ring for the theory $\mathcal{D}_{6}(SU(5), [4,1])$ using the conjecture in Section \ref{sec:conj} is the same as the bifiltered ring for the $(A_1, A_9)$ SCFT, as worked out by syzygy maximization.

\subsection{Higgsing \texorpdfstring{$\mathcal{D}_{N+1}(SU(N))$}{DN+1(SU(N))} to \texorpdfstring{$\mathcal{D}_{N+1}(SU(N), [N])$}{DN+1(SU(N), [N])}}\label{sec:genericpNp1}

It is well-known that the $\mathcal{D}_{N+1}(SU(N))$ Argyres--Douglas theory has no non-Higgsable sector at the generic point of the Higgs branch.\footnote{The non-Higgsable Argyres--Douglas theories, studied in \cite{Giacomelli:2020ryy,Carta:2021whq,Carta:2021dyx}, for example, are a useful collection of SCFTs to further explore our conjectures on the existence and structure of the bifiltered rings, as the Higgs branches are by definition trivial.} That is, there is a Higgs branch RG flow:
\begin{equation}
    \mathcal{D}_{N+1}(SU(N)) \,\,\rightarrow\,\, \mathcal{D}_{N+1}(SU(N), [N]) = (A_0, A_{N-1}) =  \varnothing \,,
\end{equation}
where we write $\varnothing$ to denote the trivial theory. 
In the case of $p = N+1$, the syzygy-maximized universal potential, which is a degree $N + 2$ polynomial, takes a rather illuminating generic form:
\begin{equation}
    P(u_2, \cdots, u_N) = \sum_{j=0}^{\floor{\frac{N-2}{2}}} \alpha_j u_{2+j}u_{N-j} + \cdots \,,
\end{equation}
where the subscript on the $u_i$ denotes the degree, and the $\cdots$ captures higher order than quadratic terms. Let us assume that all the $\alpha_j$ are non-zero.\footnote{This follows from the general construction in \cite{SYZYGY}.} Then, for the Slodowy slice associated to the nilpotent orbit $[N]$, the affine coordinates are directly the $u_i$ in the universal polynomial, and thus the Jacobian ideal takes the form
\begin{equation}
    \left( \partial_{u_2} P, \cdots, \partial_{u_N} P \right) = \left( u_2 - \cdots, u_3 - \cdots, \cdots, u_N - \cdots \right) \,,
\end{equation}
that is, it is generated by weighted-homogeneous equations with a linear term for each of the $u_i$. We can solve for each of these using triangular elimination, and the bifiltered ring following from the conjecture in equation \eqref{eqn:conj} is then just 
\begin{equation}
    R = \mathbb{C} \,.
\end{equation}
This is expected since we know that $\mathcal{D}_{N+1}(SU(N), [N])$ is the trivial theory.

\subsection{Higgsing and isomorphisms for \texorpdfstring{$p < N$}{p < N}}\label{sec:isos}

For $p < N$, it is well-known that the Higgs branch of $\mathcal{D}_p(SU(N))$, with $\gcd(p,N) = 1$, is \emph{not} the closure of the principal nilpotent orbit of $\mathfrak{sl}(N, \mathbb{C})$. That is, it is not possible to perform standard nilpotent Higgsing (where the only operators to decouple along the flow are the Nambu--Goldstone modes coming from the moment map) for all nilpotent orbits $O$. In particular, Higgsings corresponding to integer partitions of $N$:
\begin{equation}\label{eqn:swiper}
    O \, : \quad [N^{m_N}, \cdots, 1^{m_1}] \,,
\end{equation}
are only allowed if they satisfy the relation \cite{Couzens:2023kyf,Beem:2023ofp}:
\begin{equation}\label{eqn:validpart}
    \sum_{i=1}^N m_i \operatorname{min}(i,p)  = N \,.
\end{equation}

A general family of isomorphisms between $\mathcal{D}_p(SU(N), O)$ theories has been found in \cite{Xie:2019yds,Beem:2023ofp}, which we describe as follows. From $O$ as defined in equation \eqref{eqn:swiper}, we define the $p$-dependent nilpotent orbit:
\begin{equation}
    O' \,: \quad [(p-1)^{m_1}, \cdots, (p-N)^{m_N}] \,,
\end{equation}
which is automatically well-defined only when $p > N$, and for $p < N$ whenever the partition $O$ satisfies equation \eqref{eqn:validpart}.
We can see easily that this is an integer partition of
\begin{equation}
    p M - N \,, \qquad \text{where} \qquad M = \sum_{i=1}^N m_i \,.
\end{equation}
Then, \cite{Xie:2019yds,Beem:2023ofp} proposes the isomorphism:
\begin{equation}\label{eqn:DpGiso}
    \mathcal{D}_p(SU(N), O) \,\, \cong \,\, \mathcal{D}_p(SU(pM - N), O') \,.
\end{equation}
In this section, we provide some examples demonstrating how our proposed bifiltered ring in Section \ref{sec:conj} respects these isomorphisms. In each case we study, we have either $p < N$ or $p < N' = pM - N$ in equation \eqref{eqn:DpGiso}.

\paragraph{\uline{\boldmath{$\mathcal{D}_{3}(SU(4), [3,1])$}:}}

As our first example, we consider $\mathcal{D}_{3}(SU(4), [3,1])$, which is the largest nilpotent orbit of $\mathfrak{su}(4)$ (in the dominance partial ordering on nilpotent orbits) by which the theory can be Higgsed. The universal potential for the $\mathcal{D}_3(SU(4), O)$ family of theories was determined in Section \ref{sec:DpSUN}, and it was found to be
\begin{equation}\label{eqn:D3SU4Pu}
    P_u(u_2, u_3, u_4) = u_4 - \frac{3}{8} u_2^2 \,.
\end{equation}
We have $n_1, \cdots, n_5$ as the affine coordinates on the Slodowy slice associated to the $[3,1]$ nilpotent orbit. The relevant Casimir invariants associated to the Slodowy slice were given in equation \eqref{eqn:su431Cas}.
Therefore, by restricting the Casimirs that are fed into the universal potential in equation \eqref{eqn:D3SU4Pu}, we would find that the potential for $\mathcal{D}_{3}(SU(4), [3,1])$ is
\begin{equation}
    P_{[3,1]}(n_1, \cdots, n_5) = \frac{1}{2} (15 n_1^4 - 6 n_1^2 n_2 + n_2^2 + 2 n_3 n_4 - 6 n_1 n_5) \,,
\end{equation}
and thus the relevant ideal, from equation \eqref{eqn:ringgen}, (where we have chosen a minimal set of generators) for the bifiltered ring is
\begin{equation}
    I_{[3,1]} = \left( n_1, n_2, n_3, n_4, n_5, w_2, w_3, w_4 \right) \,,
\end{equation}
which is an ideal belonging to the polynomial ring
\begin{equation}
    S_{[3,1]} = \mathbb{C}[n_1, n_2, n_3, n_4, n_5, w_2, w_3, w_4] \,.
\end{equation}
Therefore, we find that the quotient ring contains no non-trivial generators:
\begin{equation}
    R_{[3,1]} = \mathbb{C} \,,
\end{equation}
which indicates that $\mathcal{D}_3(SU(4), [3,1])$ is the trivial theory. This is the same as $\mathcal{D}_3(SU(2), [2])$, which is also the trivial theory,\footnote{We observe that the conjecture in equation \eqref{eqn:conj} applied to $\mathcal{D}_3(SU(2), [2])$ also leads to $R = \mathbb{C}$.} and thus we see agreement with the proposed isomorphism in equation \eqref{eqn:DpGiso}.

\paragraph{\uline{\boldmath{$\mathcal{D}_{3}(SU(4), [2^2])$}:}}

We can also consider the theory $\mathcal{D}_3(SU(4), [2^2])$, which is expected to still be a non-trivial interacting SCFT with a Higgs branch which is $\mathbb{C}^2/\mathbb{Z}_2$. The Casimir invariants restricted to the Slodowy slice of the $[2^2]$ orbit are
\begin{equation}
  \begin{aligned}
    C_2|_{\mathcal{S}_{[2,2]}}(\bm{n}) &= 2n_1n_2 + 2n_3^2 + n_4 + n_7 \,,\\
    C_3|_{\mathcal{S}_{[2,2]}}(\bm{n}) &= 2(n_1n_6 + n_2 n_5 - n_3n_4 + n_3n_7) \,,\\
    C_4|_{\mathcal{S}_{[2,2]}}(\bm{n}) &= \frac{1}{2}\Big(
2n_1^2n_2^2 +4n_1n_2n_3^2 +6n_1n_2n_4 +6n_1n_2n_7 \\&\qquad\qquad +2n_3^4 +6n_3^2n_4 +6n_3^2n_7 +n_4^2 +2n_5n_6 +n_7^2 \Big)\,,
  \end{aligned}
\end{equation}
and the half-Kazhdan grading on the affine coordinates on the slice are
\begin{equation}
    (\operatorname{deg}(n_1), \cdots, \operatorname{deg}(n_7)) = (1,1,1,2,2,2,2) \,.
\end{equation}
A set of minimal generators of the Jacobian ideal, that is, just from taking the derivatives of $P_{[2^2]}$ with respect to the $n_i$, is then
\begin{equation}
    \left( n_6, n_5, n_4 - n_7, n_3 n_7, n_2 n_7, n_1 n_7, 3n_1 n_2 + 3n_3^2  - n_7 \right) \,.
\end{equation}
Now, we can determine the ideal $I_{[2^2]}$ which we would associate to the SCFT following equation \eqref{eqn:ringgen}, where we take the minimal set of generators that are consistent with the bifiltration:
\begin{equation}
  \begin{aligned}
    I_{[2^2]} &=
    \big( n_1 w_2, n_2 w_2, n_3 w_2, w_2 - (n_1 n_2 + n_3^2), w_4 - w_2^2\big) \\
    &=
    \big( n_1 w_2, n_2 w_2, n_3 w_2, w_2 - (n_1 n_2 + n_3^2)\big)\,,
  \end{aligned}
\end{equation}
where, in the last line, we have used the fact that $w_4$ has a larger $T$-grading than $w_2^2$ to solve the linear equation for $w_4$ without modifying the bifiltration. The final ideal
belongs to the polynomial ring:
\begin{equation}\label{eqn:dora}
    \mathbb{C}[n_1, n_2, n_3, w_2] \,, \qquad \operatorname{deg}(n_i) = (1,1) \,, \operatorname{deg}(w_2) = (2,1) \,.
\end{equation}
Therefore, we have discovered that the bifiltered ring for $\mathcal{D}_{3}(SU(4), [2^2])$ is identical to the bifiltered ring for $\mathcal{D}_{3}(SU(2)) = (A_1, A_3)$, determined in \cite{Kang:2025zub,Andrews:2025krn}, or, equivalently, that there exists a Higgs branch RG flow:
\begin{equation}
    \mathcal{D}_3(SU(4)) \quad \longrightarrow \quad \mathcal{D}_3(SU(2)) \,.
\end{equation}
This is as expected from the isomorphism of theories in equation \eqref{eqn:DpGiso}.

\paragraph{\uline{\boldmath{$\mathcal{D}_{3}(SU(5), [3, 1^2])$}:}} In a similar vein, we consider the nilpotent Higgsing of the $\mathcal{D}_3(SU(5))$ Argyres--Douglas theory. We consider the Higgsing with respect to the $[3, 1^2]$ nilpotent orbit, which is the second-largest orbit with which we can Higgs the theory.\footnote{The largest allowed orbit is the $[3,2]$, which leads to an empty theory, as can be seen from the bifiltered ring.} The Slodowy slice, $\mathcal{S}_{[3,1^2]}$, has ten affine coordinates with half-Kazhdan degrees:
\begin{equation}
    (\operatorname{deg}(n_1), \cdots, \operatorname{deg}(n_{10})) = (1,1,1,1,2,2,2,2,2,3) \,.
\end{equation}
We do not explicitly write the restrictions of the four Casimirs to the Slodowy slice here, but we can determine the ideal, using the general conjecture in equation \eqref{eqn:ringgen}. We find
\begin{equation}
  \begin{aligned}
    I_{[3,1^2]} &=
    \big( n_1 w_2, n_2 w_2, n_3 w_2, w_2 - (n_1 n_2 + n_3^2)\big)\,,
  \end{aligned}
\end{equation}
exactly the same as in the $\mathcal{D}_3(SU(4), [2^2])$ example, and with coordinates belonging to the same polynomial ring as given in equation \eqref{eqn:dora}. Therefore, the bifiltered ring is showcasing a Higgs branch RG flow:
\begin{equation}
    \mathcal{D}_3(SU(5)) \quad \longrightarrow \quad \mathcal{D}_3(SU(2)) \,.
\end{equation}

\paragraph{\uline{\boldmath{$\mathcal{D}_{4}(SU(5), [3, 2])$}:}} As a final example, we consider an isomorphism of theories with a non-maximal nilpotent orbit on both sides of equation \eqref{eqn:DpGiso}:
\begin{equation}
    \mathcal{D}_4(SU(5), [3,2]) \cong \mathcal{D}_4(SU(3), [2,1]) \,.
\end{equation}
For $\mathcal{D}_4(SU(5), O)$, the universal potential was 
\begin{equation}
    P_u(u_2, u_3, u_4, u_5) = u_5 -\frac{4}{5}u_2 u_3 \,,
\end{equation}
and the $[3,2]$ nilpotent orbit has an associated Slodowy slice with eight affine coordinates with half-Kazhdan degrees
\begin{equation}
    (\operatorname{deg}(n_1), \cdots, \operatorname{deg}(n_{8})) = \left( 1, \frac{3}{2}, \frac{3}{2}, 2, 2, \frac{5}{2}, \frac{5}{2}, 3 \right) \,.
\end{equation}
The Jacobian ideal has homogeneous linear relations for $n_5, \cdots, n_8$, and thus we can solve for those coordinates to obtain the ideal
\begin{equation}\label{eqn:halfideal}
    \left( n_2 n_4, n_3 n_4, n_2 n_3 + n_1^3 + 2n_1 n_4, n_4(n_4 + 3 n_1^2)  \right) \,.
\end{equation}
Next, we consider the generators of the ideal in equation \eqref{eqn:ringgen} that are linear in the $w_i$. Up to terms in the ideal in equation \eqref{eqn:halfideal}, we can see that these can be written as
\begin{equation}
    \left(w_2 - n_4, w_3 - n_1 n_4, w_4 - n_4^2, w_5 - n_1 n_4^2 \right) \,.
\end{equation}
Based on the $T$-degrees, we can solve the linear conditions for $n_4$, $w_4$, and $w_5$, and therefore we end up with the bifiltered ring formed from the polynomial ring:
\begin{equation}
  \begin{gathered}
    \mathbb{C}[n_1, n_2, n_3, w_2, w_3] \\
    \operatorname{deg}(n_1) = (1,1) \,, \quad \operatorname{deg}(n_2) = \operatorname{deg}(n_3) = \left(\frac{3}{2},\frac{3}{2}\right) \,,\\ \operatorname{deg}(w_2) = (2,1) \,, \quad\operatorname{deg}(w_3) = (3,2) \,,
  \end{gathered}
\end{equation}
and the ideal:
\begin{equation}\label{eqn:diego}
    \left(  n_2 w_2, n_3 w_2, n_2 n_3 + n_1^3 + 2n_1 w_2, w_2(w_2 + 3 n_1^2), w_3 - n_1 w_2\right)\,.
\end{equation}

Next, we turn to the proposed isomorphic theory: $\mathcal{D}_4(SU(3), [2,1])$. The universal potential is simply 
\begin{equation}
    P_u(u_2, u_3) = u_2 u_3 \,.
\end{equation}
On the $[2,1]$ Slodowy slice, there are four affine coordinates, and the restrictions of the two Casimirs to the slice were already given in equation \eqref{eqn:thomas}. The half-Kazhdan degrees of these coordinates are
\begin{equation}
    (\operatorname{deg}(n_1), \cdots, \operatorname{deg}(n_4)) = \left( 1, \frac{3}{2}, \frac{3}{2}, 2 \right) \,.
\end{equation}
The relevant ideal is then
\begin{equation}
    \left(n_2 w_2, n_3 w_2, n_2 n_3 + n_1^3 + 2n_1 w_2, w_2(w_2 + 3 n_1^2), w_3 - (n_2 n_3 + n_1^3 + n_1 w_2) \right) \,,
\end{equation}
where we have solved for the linear relation $w_2 - n_4$ using the fact that the $T$-degree of $w_2$ is less than that of $n_4$. This ideal is identical as for the bifiltered ring derived from syzygy maximization in Section \ref{sec:RINGS}, as expected. It is also straightforward to see that it is isomorphic as a bifiltered ring to the $\mathcal{D}_{4}(SU(5), [3,2])$ ring that we derived from the ideal in equation \eqref{eqn:diego}. Again, we see that our bifiltered rings respect the isomorphisms in equation \eqref{eqn:DpGiso}.

\paragraph{\uline{Over-Higgsing: \boldmath{$\mathcal{D}_{3}(SU(4), [4])$}:}} As mentioned previously, when $p < N$, Higgsing associated to integer partitions that do not satisfy equation \eqref{eqn:validpart} are obstructed. This is reflected in the conjectural formula in equation \eqref{eqn:conj} for the bifiltered ring underlying the Higgs branch and the Macdonald index under Higgsing. We now look at one explicit example, the SCFT $\mathcal{D}_3(SU(4))$ where the moment map of the $\mathfrak{su}(4)$ is putatively Higgsed by the VEV in the regular nilpotent orbit. The Slodowy slice associated to the regular nilpotent orbit has three coordinates, $n_1$, $n_2$, and $n_3$, and the Casimir invariants of $\mathfrak{su}(4)$ restricted to this slice are just these coordinates themselves:
\begin{equation}
    C_k|_{\mathcal{S}_{[4]}}(\bm{n}) = n_{k-1} \,.
\end{equation}
Therefore, the potential takes the form
\begin{equation}
    P_{[4]}(n_1, n_2, n_3) = n_3 - \frac{3}{8}n_1^2 \,.
\end{equation}
Since the potential is linear in $n_3$, the Jacobian ideal contains the identity as a generator, and thus the quotient ring is the zero-ring; the Hilbert series of the zero-ring vanishes, and it is clear that this cannot correspond to a quantum field theory. In our examples, this is a general feature when Higgsing by a nilpotent orbit where the underlying integer partition does not satisfy the condition in equation \eqref{eqn:validpart}, and we can treat it as an alternative diagnostic for when a nilpotent VEV does not trigger a Higgs branch RG flow to an interacting SCFT.

\section{Discussion}\label{sec:DISC}

In this paper, we have extended the algebro-geometric reconstruction of the Higgs branch and the Macdonald index from a bifiltered ring to a broad class of 4d $\mathcal N=2$ SCFTs with non-trivial Higgs branches. For the $\mathcal {D}_p(SU(N),O)$ theories, we have proposed that the relevant bifiltered ring is obtained from a universal weighted-homogeneous potential $P_u$, depending only on $p$ and $N$, by pulling the Casimir invariants back to the Slodowy slice associated to the nilpotent orbit $O$. In the examples studied explicitly, the coefficients of the potential are fixed by syzygy maximization, similarly to the theories with point-like Higgs branches in \cite{Kang:2026nge}, and the resulting rings reproduce the known indices, or their Schur and Hall--Littlewood limits, while their reduced spectra reproduce the corresponding Higgs branches.

We have furthermore demonstrated that these rings transform naturally under nilpotent Higgs branch RG flows. The universal potential is preserved, while the Casimir invariants are restricted to the appropriate Slodowy slice and the generators are assigned the half-Kazhdan grading in the infrared. This prescription reproduces the bifiltered rings of the interacting (or free) infrared fixed points, respects non-trivial isomorphisms between different $\mathcal{D}_p(SU(N),O)$ presentations, and even detects obstructed Higgsings through the appearance of the zero ring.

Beyond (nilpotent) Higgs branch RG flows, there are two further natural operations on 4d $\mathcal{N}=2$ SCFTs: mass deformations and the gauging of global symmetries,\footnote{Here, we restrict our attention to the gauging of continuous global symmetries. It would also be interesting to understand how the gauging of discrete global symmetries, particularly those acting non-trivially on the Higgs branch, is reflected in the bifiltered rings \cite{Argyres:2017tmj,Argyres:2018wxu,Bourget:2018ond,Bourton:2018jwb,Giacomelli:2024sex,Lawrie:2025exx,DISCRETEEN}.} and it is natural to ask how each of these operations is reflected at the level of the bifiltered rings. We discuss each of these operations briefly here, in Sections \ref{sec:gauging} and \ref{sec:mass}. Finally, we discuss whether the rings admit a direct construction from a six-dimensional SCFT origin in Section \ref{sec:6d}, and whether magnetic quivers for the Higgs branches can be augmented so as to encode the non-reduced structure invisible to the ordinary Higgs branch in Section \ref{sec:MQ}.

\subsection{Gauging}\label{sec:gauging}

Let $\mathcal{T}_i$, for $i = 1, \cdots, n$, denote a collection of 4d $\mathcal{N}=2$ SCFTs with a global symmetry which contains a simple Lie algebra $\mathfrak{g}$. Let $k_i$ denote the flavor central charge of the $\mathfrak{g}$ in the theory, $\mathcal{T}_i$. If the condition that 
\begin{equation}
    \sum_i k_i = 4 h_\mathfrak{g}^\vee \,,
\end{equation}
is satisfied, and modulo any obstructing anomalies, then we can consider a new 4d $\mathcal{N}=2$ SCFT via the diagonal gauging of each of the $\mathfrak{g}$ flavor symmetries. Let $G$ denote the global form of $\mathfrak{g}$ that we gauge, and then we can write the gauging schematically as
\begin{equation}
    \mathcal{T} = (\mathcal{T}_1 \times \cdots \times \mathcal{T}_n) \, / \, G \,.
\end{equation}
It is well-known how both the Higgs branch and the Macdonald index are modified under such a conformal gauging. The Higgs branch after gauging is the hyperk\"ahler quotient of the product of the Higgs branches of the individual $\mathcal{T}_i$:
\begin{equation}
    \mathcal{H}(\mathcal{T}) = \left(\, \prod_{i=1}^n \mathcal{H}(\mathcal{T}_i) \, \right)\, ///\,\, G \,.
\end{equation}
For the Macdonald index, we simply integrate over the fugacities, $\bm{z}$, associated to the $\mathfrak{g}$ global symmetries, together with the inclusion of the index for the $\mathcal{N}=2$ vector multiplet for the gauge group $G$:
\begin{equation}
    I_\text{Mac}^{\mathcal{T}}(q,T) = \int [d\bm{z}] I_\text{Mac}^{\mathcal{N}=2 \text{ vec, }G}(q,T,\bm{z}) \prod_{i=1}^n I_\text{Mac}^{\mathcal{T}_i}(q,T,\bm{z}) \,.
\end{equation}
Here, $[d\bm{z}]$ is the Haar measure for $G$, and we have ignored any fugacities associated to other global symmetries. Since the bifiltered ring is proposed to encode both of these quantities, it is natural to seek an operation on bifiltered rings that simultaneously lifts the hyperk\"ahler quotient and the Macdonald index gauging prescription. 

At the level of the associated VOA, conformal gauging is implemented by taking a relative BRST cohomology after adjoining an adjoint-valued $bc$-ghost system \cite{Beem:2013sza,Beem:2014rza}. It is thus reasonable to conjecture that gauging is realized on the bifiltered rings by an appropriate bifiltered BRST reduction whose reduced spectrum reproduces the hyperk\"ahler quotient and whose arc-space Hilbert series reproduces the Macdonald-index gauging prescription.

For the $\mathcal{D}_p(SU(N))$ theories, trivalent and quadrivalent $\mathcal{N}=2$ conformal gaugings of $\mathfrak{su}(N)$ flavor symmetries have been discussed, see, for example, \cite{Kang:2022zsl,Kang:2021lic,Buican:2020moo,Kang:2022vab,Closset:2020afy,Cecotti:2013lda}. For example, one can gauge together three copies of $\mathcal{D}_3(SU(N))$, with $\gcd(3,N)=1$; the resulting SCFTs were called $\widehat{E}_6(SU(N))$ in \cite{Kang:2021lic}, and furthermore their Higgs branches are pointlike. In Section \ref{sec:RINGS}, we have determined the bifiltered rings associated to the $\mathcal{D}_3(SU(4))$ and $\mathcal{D}_3(SU(5))$ SCFTs, and we also know the ring associated to $\mathcal{D}_3(SU(2))$ from \cite{Kang:2025zub,Andrews:2025krn}. This makes this class of theories a natural starting point for studying the behavior of the bifiltered rings under gauging; some such examples are discussed in detail in \cite{SUPERPOINT}.

Generalizations of Argyres--Seiberg duality \cite{Argyres:2007cn} involving gauging of Argyres--Douglas SCFTs were constructed in \cite{Buican:2014hfa}. These theories are obtained by conformally gauging diagonal $\mathfrak{su}(2)$ or $\mathfrak{su}(3)$ flavor symmetries of rank-one Argyres--Douglas sectors, together with additional hypermultiplets. The different weakly-coupled cusps of their conformal manifolds provide further non-trivial tests of the conjectural bifiltered BRST reduction: applying the reduction in the different duality frames should produce isomorphic bifiltered rings.

Going beyond the Argyres--Douglas SCFTs, an obvious arena in which to discuss gauging is class $\mathcal{S}$ \cite{Gaiotto:2009we,Gaiotto:2009hg}. This construction involves the compactification of the 6d $(2,0)$ SCFT of type $\mathfrak{g}$ , where $\mathfrak{g}$ is a simple and simply-laced Lie algebra, on an arbitrary genus Riemann surface with arbitrary regular punctures.\footnote{There is a distinction between untwisted and twisted regular punctures: here we will only consider untwisted punctures. See \cite{Couzens:2023kyf} for a recent summary and our conventions.} Each untwisted regular puncture is labeled by a nilpotent orbit of $\mathfrak{g}$ \cite{Chacaltana:2012zy}, and we can schematically write such a class $\mathcal{S}$ theory as:
\begin{equation}
    \mathcal{S}_{\mathfrak{g}}\langle C_{g,n} \rangle \{ O_1, \cdots, O_n \} \,.
\end{equation}
A stable $n$-punctured, genus-$g$ Riemann surface has a pair-of-pants decomposition in terms of three-punctured spheres; accordingly, we can first consider class $\mathcal{S}$ theories, called fixtures, obtained by compactification on three-punctured spheres \cite{Chacaltana:2010ks}. If we consider $\mathfrak{g} = \mathfrak{su}(N)$, then each three-punctured sphere can be obtained by starting from the trinion theory
\begin{equation}
    T_N \,:\quad \mathcal{S}_{\mathfrak{su}(N)} \langle C_{0,3} \rangle \{ [1^N], [1^N], [1^N] \} \,.
\end{equation}
Such theories have a manifest $\mathfrak{su}(N)^{\oplus 3}$ flavor symmetry, and any regular $A_{N-1}$ fixture can be obtained by nilpotent Higgsing of the moment maps of each of these three factors. Thus, if the Slodowy slice prescription of Section \ref{sec:conj} extends to general nilpotent Higgsings, then, if we know the bifiltered ring for $T_N$, we are proposing the bifiltered ring for any three-punctured sphere with $\mathfrak{g}=\mathfrak{su}(N)$. The pair-of-pants decomposition of a generic Riemann surface is interpreted physically as the gauging of $\mathfrak{su}(N)$ global symmetries associated to $[1^N]$ punctures, and different decompositions correspond to different S-duality frames. Therefore, if both the bifiltered ring associated to $T_N$ and the conjectural bifiltered BRST reduction were known, then the bifiltered ring of any untwisted class $\mathcal S$ theory of type $A_{N-1}$ could, in principle, be reconstructed from a pair-of-pants decomposition. Independence of the resulting ring from the choice of pair-of-pants decomposition would then encode the corresponding S-dualities as non-trivial isomorphisms of bifiltered rings.

The aforementioned Argyres--Seiberg duality provides a particularly sharp test of such a gauging operation. In class $\mathcal S$, distinct pair-of-pants decompositions of the same punctured Riemann surface give different weakly-coupled presentations of the same SCFT. The canonical example identifies $\mathfrak{su}(3)$ gauge theory with six fundamental hypermultiplets with an $\mathfrak{su}(2) \subset \mathfrak{e}_6$ gauging of the rank-one $E_6$ Minahan--Nemeschansky SCFT \cite{Minahan:1996fg} coupled to a fundamental hypermultiplet \cite{Argyres:2007cn,Gaiotto:2009we}. The conjectural
BRST-type reduction of the bifiltered rings associated to these two matter systems should therefore produce isomorphic bifiltered rings. This would lift the known equivalence of the corresponding hyperk\"ahler quotients \cite{Gaiotto:2009jjh} to a non-trivial isomorphism of bifiltered rings, simultaneously encoding the equality of their Higgs branches and Macdonald indices.  More generally, in \cite{Distler:2022kjb}, there is a large collection of class $\mathcal{S}$ theories which are identical as physical systems, but which do not correspond to dualities in the sense of different degeneration limits of the punctured Riemann surface. Such equivalences would provide an even stronger test: the associated bifiltered rings should be isomorphic despite the absence of an ordinary S-duality relation generated by different pair-of-pants decompositions. 

\subsection{More general RG flows}\label{sec:mass}

We now turn to a more general class of RG flows between 4d $\mathcal{N}=2$ SCFTs, beyond Higgs branch RG flows triggered by giving a nilpotent vacuum expectation value to a moment map.\footnote{In this section, we mean ``mass deformation'' in a rather broad sense: either a mass deformation, together with a choice of vacuum that we expand around in the infrared; a mass deformation together with a chiral deformation \cite{Argyres:2015ffa}; or a SUSY-enhancing $\mathcal{N}=1$ deformation (flipped nilpotent Higgsing) \cite{Maruyoshi:2016aim, Agarwal:2016pjo}.} Unlike a Higgs branch RG flow, a mass deformation is not naturally described by restricting to a transverse slice of the Higgs branch. Instead, it deforms the Coulomb branch singularity structure, and interacting infrared SCFTs arise at distinguished singular loci of the deformed Coulomb branch. It is therefore not immediately obvious what operation a mass deformation should induce on the bifiltered ring.

As an example, we have the following infinite family of deformations for which we know, or have conjectured, the bifiltered rings for all ranks. The sequence is \cite{Maruyoshi:2016aim, Agarwal:2016pjo, Giacomelli:2025zqn}
\begin{equation}\label{eqn:oksusu}
    \mathcal{D}_2(SU(2n+1)) \quad \rightarrow \quad (A_1, D_{2n + 1}) \quad \rightarrow \quad (A_1, A_{2n}) \,.
\end{equation}
Since none of these theories requires syzygy maximization to fix the coefficients, we can write the sequence of bifiltered rings simply via the potentials:
\begin{equation}
    d_{ABC} \,a_A a_B a_C \quad \rightarrow \quad (xy + z^2)^{n+1} \quad \rightarrow \quad w^{n+2} \,.
\end{equation}
Here we have written $d_{ABC}$, which is the symmetric invariant tensor of $SU(2n+1)$, explicitly, rather than $C_3(\bm{a})$. In each case, the universal potential is a polynomial in a single invariant generator $u_i$ of bidegree $(i,i)$. Then the sequence of mass deformations changes the universal potentials as
\begin{equation}\label{eqn:blah}
    u_3 \quad \rightarrow \quad u_2^{n+1} \quad \rightarrow \quad u_2^{n+2} \,.
\end{equation}

We now provide a brief explanation of the structure appearing in equation \eqref{eqn:blah}. Let $M$ be a generic $\mathfrak{su}(2n+1)$ matrix. For the $\mathcal{D}_2(SU(2n+1))$ theory, we are considering the Jacobian ideal, schematically written as,
\begin{equation}
    \operatorname{Jac}(C_3(M)) \,.
\end{equation}
This Jacobian ideal is generated by the matrix relation
\begin{equation}\label{eqn:macademia}
    M^2 - \frac{1}{2n+1} \operatorname{tr} M^2 \, \mathbf{1}_{2n+1} = 0 \,.
\end{equation}
Furthermore, let us define a coordinate $u$ as the quadratic Casimir:
\begin{equation}
    u = \frac{1}{2} \operatorname{tr} M^2 \,.
\end{equation}
Combining the Jacobian relation in equation \eqref{eqn:macademia} with the Cayley–Hamilton identity for the $(2n+1) \times (2n+1)$ matrix $M$ tells us that the Jacobian ideal also contains the relations
\begin{equation}\label{eqn:CH}
    u^n M = 0 \quad \text{ and } \quad u^{n+1} = 0 \,.
\end{equation}
For $\mathcal{D}_{2n+1}(SU(2)) = (A_1, D_{2n+1})$, we let $A$ be a generic matrix belonging to $\mathfrak{su}(2)$, and then we are considering the Jacobian ideal
\begin{equation}
    \operatorname{Jac}(Q^{n+1}) \,,
\end{equation}
where we have defined $Q = C_2(A)$, and which is generated by
\begin{equation}
    Q^n A = 0 \quad \Rightarrow \quad Q^{n+1} = 0 \,.
\end{equation}
This particular tuned mass deformation isolates an interacting infrared sector with $\mathfrak{su}(2)$ flavor symmetry.
This motivates the replacement, across the mass deformation, of
\begin{equation}
    M \rightarrow A \quad \text{ and } \quad u \rightarrow Q \,,
\end{equation}
and it is interesting to note that the Cayley--Hamilton consequences in equation \eqref{eqn:CH} turn into the relations from $\operatorname{Jac}(C_2(A)^{n+1})$. In particular, the exponent $n+1$ in the middle potential of equation \eqref{eqn:blah} is expected from this perspective. This matching does not specify a homomorphism between the two Jacobian rings. In particular, equation \eqref{eqn:CH} is a higher consequence of the ultraviolet Jacobian equations, whereas $Q^n A = 0$ generates the infrared Jacobian ideal; it remains to determine what happens to the relation in equation \eqref{eqn:macademia} under the mass deformation. This requires more precise tracking of the explicit masses turned on in this flow: the full mass deformation operation involves tuning to the relevant Coulomb branch singularity and the subsequent infrared scaling limit, rather than just restricting the ultraviolet ring. Furthermore, the second arrow in equation \eqref{eqn:blah} admits a more direct ring interpretation. The $\mathfrak{su}(2)$-invariant subring of the quotient by the ideal $\operatorname{Jac}(Q^{n+1})$ is
\begin{equation}
    \mathbb{C}[Q] \, / \, (Q^{n+1}) \,.
\end{equation}
After relabeling $Q$ as $w$, this is precisely the Jacobian algebra associated to the potential $w^{n+2}$ of the $(A_1, A_{2n})$ theory.
While this is evocative, we leave it for future work to fully elucidate the effect of mass deformations on the bifiltered rings, and to determine if the structure sketched here extends to the generic case.

On the other hand, an independent clue for understanding  deformations is provided by the generalized Schur partition function, $\widehat{Z}_{\mathcal{T}}(q, \alpha)$, known for certain 4d $\mathcal{N}=2$ SCFTs, $\mathcal{T}$ \cite{Deb:2025ypl, Chandra:2025qpv}.\footnote{The generalized Schur partition function can be identified with the trace of higher-power of the monodromy operator, which is computed from the BPS spectrum of massive charged particles in the Coulomb branch \cite{Cordova:2015nma, Cecotti:2015lab, Kim:2024dxu}.}
It matches the Schur index in the $\alpha = 1$ limit:
\begin{equation}
   \widehat{Z}_{\mathcal{T}}(q, 1) = I_S^\mathcal{T}(q) \,,
\end{equation}
and, for certain RG flows, distinguished values of $\alpha$ reproduce the Schur indices of the interacting infrared fixed points after deformation. The sequence of deformations in equation \eqref{eqn:oksusu} has been studied from the perspective of this generalized partition function, and the following specializations were found:
\begin{equation}
    \widehat{Z}_{\mathcal{D}_2(SU(2n+1))}(q, \alpha) = \begin{cases}
        \operatorname{HS}_{q,q,1}(\operatorname{gr}(J_\infty(R_{\mathcal{D}_2(SU(2n+1))}))) \quad &\text{ when } \quad \alpha = 1 \,, \\
        \operatorname{HS}_{q,q,1}(\operatorname{gr}(J_\infty(R_{(A_1, D_{2n + 1})}))) \quad &\text{ when } \quad \alpha = \frac{2}{2n+1} \,, \\
        \operatorname{HS}_{q,q,1}(\operatorname{gr}(J_\infty(R_{(A_1, A_{2n})}))) \quad &\text{ when } \quad \alpha = \frac{2}{2n+3} \,.
    \end{cases}
\end{equation}
These identities motivate the search for an $\alpha$-dependent algebraic/homological object whose graded character is $\widehat{Z}_{\mathcal{D}_2(SU(2n+1))}(q, \alpha)$, and for which the specializations above reproduce the Hilbert series associated to the bifiltered rings.\footnote{Since the generalized partition function does not retain the Macdonald fugacity, it is not clear whether such an object should be sensitive to the full bifiltration.}
For generic $\alpha$ the coefficients of the generalized partition function need not be non-negative integers, therefore it cannot be written as the Hilbert series of a graded ring.

\subsection{Six-dimensional origins of the bifiltered rings}\label{sec:6d}

The discussion of mass deformations in Section \ref{sec:mass} in fact raises a
deeper question: can the bifiltered rings associated to 4d $\mathcal{N}=2$
SCFTs themselves be understood as descending, after compactification, from the
physics of a 6d parent theory. There is by now substantial evidence that
compactifications of 6d $(1,0)$ SCFTs provide a comprehensive organizational
principle for large classes of known 4d $\mathcal{N}=2$ SCFTs
\cite{Giacomelli:2025zqn,Giacomelli:2024ycb,Giacomelli:2024dbd,Distler:2022kjb}. If we believe
that all information about the 4d theory should be encoded in the 6d parent, we
are on the hook to explain the ultimate origin of the bifiltered ring in six-dimensions.

As an example, consider the torus compactifications of the exceptional rank-one conformal matter theories; each resulting 4d $\mathcal{N}=2$ SCFT is dual to a class $\mathcal{S}$ theory of exceptional type, as verified in detail in \cite{Baume:2021qho}. The Higgs branch of each 6d SCFT is identical to the Higgs branch of the resulting class $\mathcal{S}$ theory. Thus, we can consider the Higgs branch, as a variety, is intrinsically inherited from 6d. Where might the 4d scheme-ification of the Higgs branch, capturing the OPE decoupling relations, come from in this higher-dimensional picture? A key feature of all 6d SCFTs is that they describe the dynamics of tensionless BPS strings.
It is tempting to speculate that the scheme structure comes from properties of these intrinsically strongly-coupled string-like excitations.

Intriguingly, in \cite{DelZotto:2016pvm}, the authors study the 4d $\mathcal{N}=2$ SCFT $H_2$, which is identical to the $\mathcal{D}_2(SU(3))$ SCFT studied here; a certain twisted-compactification of $H_2$ is the same as the worldvolume theory on the BPS strings of a certain 6d $(1,0)$ SCFT. From this approach, it was shown in \cite{DelZotto:2016pvm} that there exists a two-fugacity extension of the Schur index of $\mathcal{D}_2(SU(3))$ which is given in terms of the elliptic genera of the aforementioned BPS strings. One limit of this two-fugacity index recovers the Schur index, whereas another limit recovers the Higgs branch Hilbert series. While this is highly evocative of $J_\infty(R_S)$ with its $(q,p)$-bigrading, since $J_\infty(R_S)$ also encodes the Higgs branch Hilbert series, this two-fugacity Schur index does not have an obvious interpretation as the Hilbert series of the arc space,
\begin{equation}
    H_{p,q}(J_\infty(R_S)) \,,
\end{equation}
of the $q$-graded ring $R_S$ discussed here, where the jet-grading is not specialized to $q$. It would be interesting to understand to what extent the 4d Schur sector, and in particular the additional structure captured by the bifiltration considered here, can be reconstructed directly from the tensionless BPS-string sector of an engineering 6d $(1,0)$ SCFT.

More generally, start with the 6d $(1,0)$ SCFTs known as the rank $N$
$(\mathfrak{e}_8, \mathfrak{g})$ Higgsed orbi-instantons
\cite{DelZotto:2014hpa},\footnote{This class of theories has been well-studied,
see \cite{Lawrie:2023uiu} for a recent short
review.} compactified on a torus (possibly with Stiefel--Whitney
\cite{Ohmori:2018ona,Giacomelli:2020jel,Giacomelli:2020gee,Heckman:2022suy,Giacomelli:2024dbd}
or outer-automorphism \cite{Tachikawa:2010vg} twists), and then perform a
sequence of mass deformations. In \cite{Giacomelli:2025zqn,Giacomelli:2024ycb},
it has been shown that all class $\mathcal{S}$ theories with $\mathfrak{g} =
\mathfrak{su}(N)$ and regular untwisted punctures, as well as all
$\mathcal{D}_p(SU(N), O)$ SCFTs can be obtained from such 6d origins. Consequently, a six-dimensional construction of the bifiltered ring for the
orbi-instanton theories, together with an understanding of how this structure
descends under compactification and behaves under the mass deformations of
Section \ref{sec:mass}, would provide a common origin for the bifiltered rings
of this large family of four-dimensional theories.

\subsection{Magnetic quivers and bifiltered affine schemes}\label{sec:MQ}

Each bifiltered ring that we write down defines a bifiltered affine scheme, $X$, via the standard machinery of algebraic geometry. The Higgs branch moduli space of supersymmetric vacua of the 4d $\mathcal{N}=2$ SCFT is obtained by passing to the reduced subscheme $X_\text{red}$. One powerful approach to determine properties of the Higgs branch physics is the technique of ``magnetic quivers'' \cite{Hanany:1997gh,Hanany:1996ie,Ferlito:2017xdq}. In particular, if one can find a 3d $\mathcal{N}=4$ Lagrangian theory, $\mathcal{T}_M$, the magnetic quiver, such that its Coulomb branch is isomorphic to the Higgs branch of the 4d $\mathcal{N}=2$ SCFT, $\mathcal{T}$, in question:\footnote{In fact, the Higgs branch can be isomorphic to a union of Coulomb branches of magnetic quivers.}
\begin{equation}\label{eqn:MQstd}
    \mathcal{C}(\mathcal{T}_M) = \mathcal{H}(\mathcal{T}) \,,
\end{equation}
then one can use the machinery for studying \emph{Lagrangian} quivers to determine the Higgs branch physics of the (non-Lagrangian) 4d $\mathcal{N}=2$ SCFT. For example: quiver subtraction \cite{Cabrera:2018ann} or the decay and fission algorithm \cite{Bourget:2023dkj,Bourget:2024mgn,Lawrie:2024wan} applied to the magnetic quiver can be used to ascertain the structure of the Higgs branch as a symplectic singularity. The quiver $\mathcal{T}_M$ is frequently called a magnetic quiver for the Higgs branch of $\mathcal{T}$. 

For each of the Argyres--Douglas theories studied in this paper, a magnetic quiver for the Higgs branch is known \cite{Giacomelli:2020ryy}. The standard magnetic quiver correspondence, in equation \eqref{eqn:MQstd}, only captures that reduced affine scheme:
\begin{equation}
    \mathcal{C}(\mathcal{T}_M) = \operatorname{Spec}(R_\mathcal{T})_\text{red} \,.
\end{equation}
By contrast, the bifiltered ring $R_\mathcal{T}$ determines a generally nonreduced affine scheme, which contains strictly more information than the underlying Higgs branch as a variety. We can ask: does the magnetic quiver $\mathcal{T}_M$ itself contain the information about the nonreduced scheme $\operatorname{Spec}(R_\mathcal{T})$, or is it necessary to augment $\mathcal{T}_M$ by some additional structure to encode the nilpotent information? It is clear from the fact that various distinct $\mathcal{D}_p(SU(N))$ theories have identical magnetic quivers, but different bifiltered affine schemes, that the answer must be the latter. Concretely, one would seek an assignment
\begin{equation}\label{eqn:MQassign}
  \mathcal{T}_M \longmapsto R_M \,,
\end{equation}
from a magnetic quiver (with additional structure specified) to a bifiltered commutative ring such that
\begin{equation}  
\operatorname{Spec}(R_M)_{\mathrm{red}}
\,\,\cong\,\,\mathcal C(\mathcal{T}_M) \,,
\end{equation}  
but for which $R_M$ is not generally reduced. For a magnetic quiver associated to a 4d SCFT, $\mathcal{T}$, we want the bifiltered ring obtained in such a way to match with the bifiltered ring associated to the Macdonald index: 
\begin{equation}
R_{M}\,\,\cong\,\, R_{\mathcal T} \,.
\end{equation}
The putative mapping in equation \eqref{eqn:MQassign} is further constrained by the requirement that $R_M$ be invariant under different magnetic quiver presentations of the same 4d theory, and furthermore must transform correctly under the quiver operations that realize Higgs branch RG flows. In particular, quiver subtraction or decay and fission should lift from relations between reduced symplectic singularities to operations or relations between the corresponding nonreduced bifiltered schemes; this lifted operation should agree with the restriction of the universal potential to Slodowy slices conjectured in Section \ref{sec:conj}. Establishing the map in equation \eqref{eqn:MQassign} would extend the magnetic quiver approach from the geometry of supersymmetric vacua to a larger protected sector of higher-dimensional SCFTs. 

\subsection*{Acknowledgements}
M.~J.~K.~and C.~L. thank the Simons Center for Geometry and Physics for hospitality during the ``23rd Simons Physics Summer Workshop: Theory, Experiment and the Emerging New Physics''. 
M.~J.~K.~is supported by the U.S.~Department of Energy, Office of Science, Office of High Energy Physics, under Award Number DE-SC0010813, and the Start-up Research Grant provided by Texas A\&M University.
C.~L.~acknowledges support from DESY (Hamburg, Germany), a member of the Helmholtz Association HGF; C.~L.~also acknowledges the Deutsche Forschungsgemeinschaft under Germany's Excellence Strategy - EXC 2121 ``Quantum Universe'' - 390833306 and the Collaborative Research Center - SFB 1624 ``Higher Structures, Moduli Spaces, and Integrability'' - 506632645.
The work of J.~S.~is supported by the National Research Foundation of Korea (NRF) grants RS-2024-00405629 and RS-2026-25482546, and the KAIST-KIAS collaboration program. 
The work of J.~S.~is also supported in part by the Walter Burke Institute for Theoretical Physics and the Leinweber Forum for Theoretical Physics at Caltech and by the U.S.~Department of Energy, Office of Science, Office of High Energy Physics, under Award Number DE-SC0011632.

\bibliography{references}{}
\bibliographystyle{sortedbutpretty}
\end{document}